\documentclass[12pt]{article}
\usepackage{amsmath}
\usepackage{graphicx,psfrag,epsf}
\usepackage{enumerate}
\usepackage{natbib}
\usepackage{amssymb}
\usepackage{booktabs} 
\usepackage{amsfonts}
\usepackage{hyperref}
\usepackage{xcolor}
\usepackage{array}
\definecolor{lightgreen}{RGB}{180,220,180}
\usepackage{amsthm}
\newtheorem{assumption}{Assumption}
\usepackage{multirow}
\usepackage{subcaption}
\usepackage{tikz}
\usetikzlibrary{arrows.meta, positioning}
\usepackage[margin=1in]{geometry}
\usepackage{multirow}
\newcommand{\blind}{0}

\begin{document}

\def\spacingset#1{\renewcommand{\baselinestretch}%
{#1}\small\normalsize} \spacingset{1}


\if0\blind
{
  \title{\bf Bootstrap-based Hypothesis Test of 2D Contours using Elastic Shape Analysis}
  \author{Susan Glenn, Justin Strait, Kelly Moran, Christopher Danly\thanks{Research presented in this article was supported by the National Security Education Center (NSEC) Informational Science  and Technology Institute (ISTI) using the Laboratory Directed Research and Development program of Los Alamos National Laboratory under project number 20240479CR-IST. Approved for public release (LA-UR-25-31179).}\hspace{.2cm}\\
    Los Alamos National Laboratory\\
    and \\
    Matthew P Selwood \\
    Lawrence Livermore National Laboratory}
  \maketitle
} \fi

\if1\blind
{
  \bigskip
  \bigskip
  \bigskip
  \begin{center}
    {\LARGE\bf Bootstrap-based Hypothesis Test of 2D Contours using Elastic Shape Analysis}
\end{center}
  \medskip
} \fi

\bigskip
\begin{abstract}

Shapes of objects in images are often complex, high-dimensional, and vary in ways not captured by standard Euclidean geometry and statistics. Statistical shape analysis encompasses methods for flexible and interpretable measurement of intrinsic shape and shape variability in geometric objects. Elastic Shape Analysis (ESA) is one such method that measures shape differences between objects, represented by contours, in a way that is invariant to rotation, scale, translation, and parameterization. Although ESA is useful for quantifying shape of objects in many image applications, formal methods for statistical inference in image-based ESA remain limited. This work introduces a hypothesis test procedure based on  empirical confidence intervals for the elastic shape distance (ESD) between a proposed underlying true shape and an estimated shape. The confidence intervals are created using a bootstrap procedure for non-smooth functionals, which accounts for the non-differentiability of the ESD. The effectiveness of the method is illustrated through both numerical studies and real-world image examples from inertial confinement fusion (ICF).

\end{abstract}

\noindent%

\spacingset{1.45}
\section{Introduction}
\label{sec:intro}

Image data consists of a matrix of intensity values where each entry represents the brightness of a pixel on a grid. A large body of literature in traditional image analysis focuses on intensity-based modeling, providing insights into local image structures \citep{gon:18}. However, such approaches may not explicitly capture the underlying shape of objects with complex morphology, motivating recent work on modeling larger-scale structures. One approach to studying object shape is to construct a simpler representation for analysis \citep{bhar:20, jai:24}. A common representation is the percentile contour, an ordered set of points in the Euclidean plane obtained by thresholding image intensities. By using a contour representation, we move beyond pixel-based analysis to modeling the object as a continuous, smooth curve, more closely aligning with the underlying global structure \citep{jai:24, bry:12}. In this work, we consider the problem of shape diagnostics, namely inference on curve-based representations, with applications to images in inertial confinement fusion (ICF).

In 2024, following over 70 years of effort around the globe, the US National Ignition Facility (NIF) achieved net energy production in a controlled thermonuclear fusion reaction for the first time. A millimeter-scale capsule filled with fusion fuel is imploded using high-powered lasers that compress the fuel to thermonuclear temperatures and pressures comparable to the interior of stars. Achieving efficient fusion burn requires an extremely high degree of implosion symmetry, which remains a substantial challenge due to hydrodynamic and plasma instabilities. The physical processes governing ICF implosions occur under extreme conditions requiring specialized instrumentation, such as neutron imaging with coded apertures, to assess implosion symmetry and guide experimental tuning. Existing reconstruction techniques struggle to estimate the underlying shape of the implosion, which typically exhibits complex morphology, and lack a framework for quantifying uncertainty. To estimate and quantify uncertainty in the shape of a contour from an ICF image with a focus on symmetry, we consider methods in statistical shape analysis.

Statistical shape analysis provides tools for comparing geometric objects after removing shape-preserving transformations such as translation, rotation, and scale. Many classical approaches in this area use discrete landmark sets to represent objects, where each landmark corresponds to a specific, meaningful feature shared across all objects \citep{kendall:84,bookstein:91,dryden_mardia:16}. While these methods can be used to discretely represent contour data, they are limited by the need to assume a fixed correspondence between points on contours.
Elastic shape analysis (ESA) is a natural approach which models contours as continuous curves without assuming pointwise correspondences between curves \citep{Sri:11,kur:12, bry:12}. This is achieved through a Riemannian geometric framework, which enables comparison between shapes of curves through a metric known as elastic shape distance (ESD). This metric computes shape similarity in a way which is invariant to curve translation, rotation, scaling, and reparameterization, where reparameterization refers to the instantaneous speed at which a curve is traversed \citep{Sri:16}. ESA has proven effective for shape comparison, summarization, and modeling of planar curves derived from noisy, discretely observed data \citep{steyer:23,luo:24}.

Despite these advantages, formal statistical inference within ESA remains limited. The shape spaces induced by removing shape-preserving transformations are nonlinear and infinite-dimensional, rendering many classical Euclidean inference methods inapplicable \citep{bhar:20}. Existing work has largely focused on descriptive statistics and modeling, including principal component analysis of shapes, Gaussian-type shape models, and computation of Karcher means and covariances \citep{kur:16, Sri:10, Abb:20}. Formal inference procedures for comparing shapes in the ESA literature have tended to focus on multi-class comparisons which rely on nonparametric methods: \cite{strait:17} used permutation tests involving ESD to compare vertebrae shapes of three different mice classes, while \cite{zhang:21} presented a $k$-sample nonparametric energy test using ESD to compare mitochondrial morphology across mice classes formed by acute exercise level. Neither can be directly applied to assess whether ICF implosion shape samples match a targeted shape of interest. 
In this paper, we propose the first ESA-based one-sample test for contour data that enables such comparisons, referred to as {\em bootstrapped ESA}, or {\bf bootESA} for short. Using the ESD and a nonparametric bootstrap that accounts for the non-smooth mapping from images to shape space, we construct empirical confidence intervals and associated hypothesis tests, and demonstrate {\bf bootESA} on both simulated and experimental ICF imaging data.

The rest of this paper is organized as follows. In Section~\ref{subsec:datasetup}, we describe the data setup in the context of ICF images, though {\bf bootESA} extends well beyond this application. The current approach to estimating symmetry in ICF data is described in section~\ref{subsec:Legpoly}. Then Section~\ref{sec:esa} presents in greater detail the background information for ESA which is used in our method. This is followed by Section \ref{sec:method}, where we introduce the novel approximate bootstrap-based hypothesis testing procedure {\bf bootESA} (Section~\ref{sec:method}). This section is divided into three parts: the method description (Section~\ref{subsec:bootESAmethod}), the mathematical framework (Section~\ref{subsec:mathmaticalFramework}), and {\bf bootESA}'s application to ICF images in particular (Section~\ref{subsec:ICFmethod}). In Section~\ref{sec:simulations}, we use simulation studies to investigate our hypothesis test procedure performance in terms of Type I Error (Section~\ref{subsec:sim1}) and power (Section~\ref{subsec:sim2}). Finally, Section~\ref{sec:data} demonstrates the results of {\bf bootESA} on data images from ICF implosions. 

\section{Background}\label{sec:background}
We begin by describing the ICF image data which motivated the {\bf bootESA} method. For a detailed discussion of the method’s generalizability to other data types, see the Supplementary Material, Section~\ref{sup:KDE}. The current approach to analyzing symmetry of ICF images is described in Section~\ref{subsec:Legpoly}, followed by background on ESA in Section~\ref{sec:esa}.

\subsection{Data Setup}\label{subsec:datasetup}
\begin{figure}
\centering
    \centering
    \includegraphics[width=0.8\linewidth]{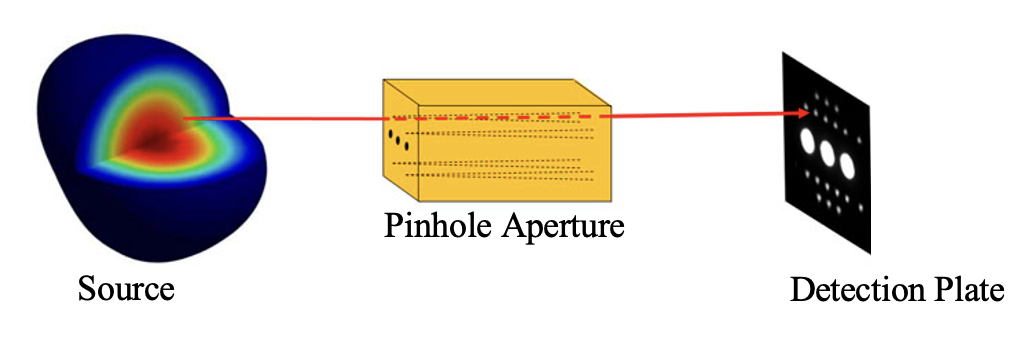}
\caption{Example of neutron source with pinhole path (red line) through pinhole aperture to detection plate adapted from \cite{lamb:22}.}
\label{fig:PinholeSetup}
\end{figure}

Each ICF image captures an implosion shot, called the source, from a single line of sight meaning the true underlying space of interest is a two-dimensional image denoted by $S$. The estimate of $S$ is obtained from neutrons emitted by the source that pass through a pinhole aperture and are recorded on a detection plate as seen in Figure~\ref{fig:PinholeSetup}. This detected image $I$ is a blurry realization of the source image $S$ where the blurring term, or {\em point spread function} (PSF) $K$, is based on pinhole location and geometry. Every pixel in the detected image is a convolution $*$ of each pixel in the source image and the PSF. The observed image $Y$ is a noisy realization of the blurry detected image due to ambient neutrons outside of the source neutrons. In most cases, each pixel in $Y$ is assumed to be sampled from a Poisson distribution centered at the corresponding pixel in $I$ \citep{lamb:22, vol:14}.

More formally, we define the source grid as $\mathcal{G}^S \subset \mathbb{Z}^2$ with each grid cell indexed by $i \in \{1,...,n_i \}$ rows and $j \in \{1,\ldots,n_j \}$ columns. The location of each grid cell in Euclidean space is defined as $X^S \in \mathbb{R}^2$ where $X^S_{ij}$ is the centroid of the $i^{\text{th}},j^{\text{th}}$ source grid cell. Following a similar notation, we define the detection grid as $\mathcal{G}^I \subset \mathbb{Z}^2$ with each grid cell indexed by $h \in \{1,...,n_h \}$ rows and $l \in \{1,\ldots,n_l \}$ columns. The location of each grid cell in Euclidean space is defined as $X^I \in \mathbb{R}^2$ where $X^I_{hl}$ is the centroid of the $h^{\text{th}},l^{\text{th}}$ data grid cell. The grids differ because $\mathcal{G}^{S}$ discretizes intrinsic geometric coordinates, while $\mathcal{G}^{I}$ discretizes the measurement space in which observations are recorded. The PSF $K_{hl,ij}$ helps navigate between the two grids  specifying how intensity from a pixel $(ij)$ in $\mathcal{G}^S$ contributes to the observed intensity at pixel $(hl)$ in $\mathcal{G}^I$.

The source intensity at each grid cell satisfies $S_{ij} \in [0,225], \forall (ij) \in \mathcal{G}^S$. The detected image intensity at pixel $(hl) \in \mathcal{G}^I$ is given by $I_{hl} = \sum_{(ij)\in\mathcal{G}^S} K_{hl,ij} S_{ij}, \forall (hl) \in \mathcal{G}^I$, and the corresponding observed image intensity is given by $Y_{hl} \sim \text{Pois}(I_{hl})$. We assume that $K$ is fixed and known (e.g., Gaussian blur), a standard simplifying assumption though the PSF is estimated with a nonstationary Gaussian operator in most experiments \citep{lamb:22}. In practice, there are multiple pinholes in the aperture, each with its own PSF, leading to multiple data realizations of the source. Let $n \in \{1,...,N\}$ index the $N$ pinholes in the aperture. The $n$th pinhole yields a detected image $I^n$ and corresponding data realization $Y^n$.

\begin{figure}
\centering
    \centering
    \includegraphics[width=0.8\linewidth]{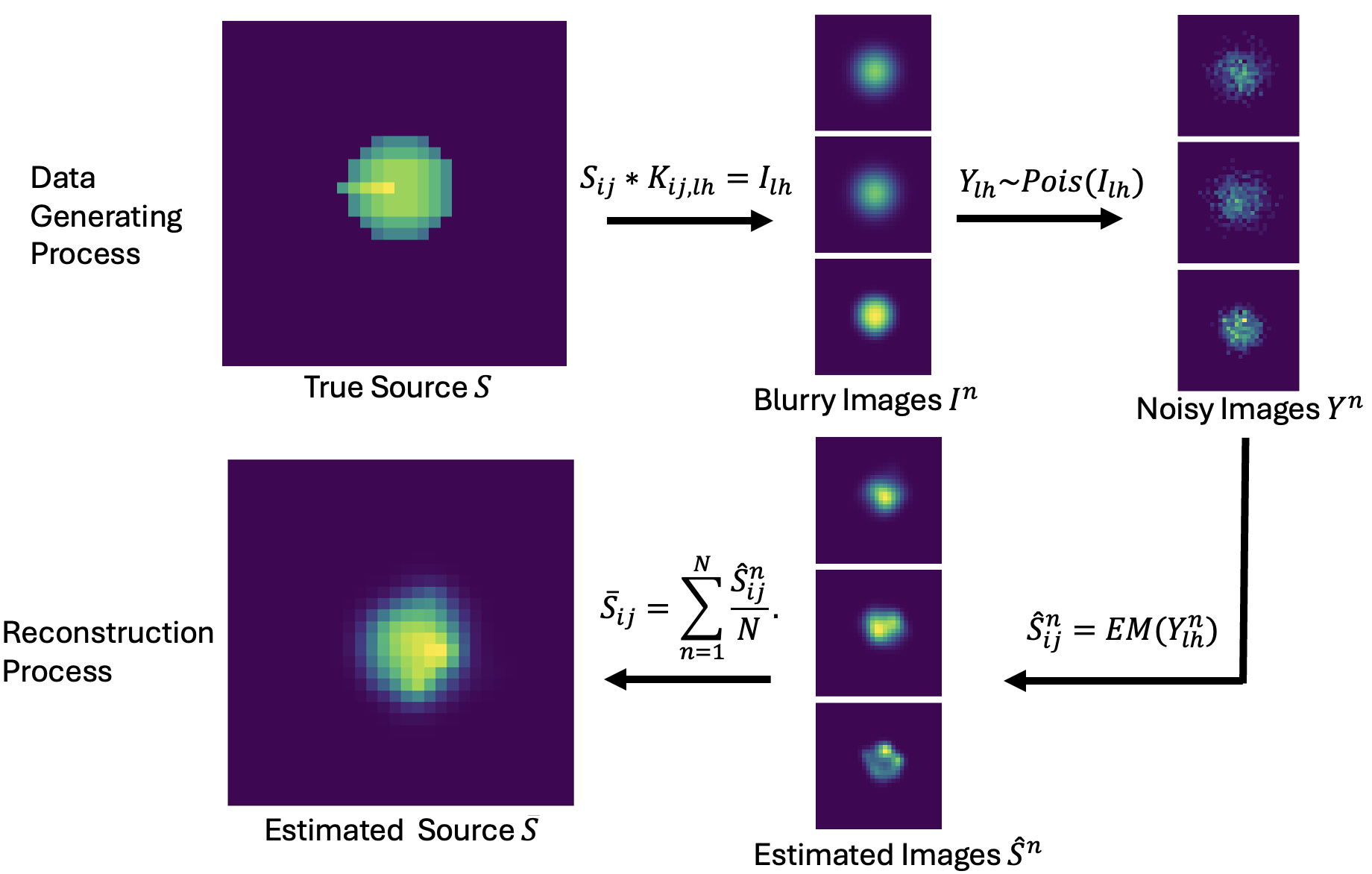}
\caption{The data generating process (first row) convolves the true source image with a PSF, yielding a blurred realization $I^n$ of the source for each pinhole $n=\{1,2,3\}$. The data images $Y^n$ are noisy realizations of $I^n$. The reconstruction process (second row) applies the EM algorithm to get $\hat S^n$ for each pinhole sub-image $Y^n$. The final source estimate is given by the sample mean image $\bar S$.}
\label{fig:two_row_images}
\end{figure}

Our objective is to estimate the shape of the implosion in $S$, the source, as opposed to the shape of the implosion in $Y^n$, a blurry, noisy realization of the source. A method outlined in \citet{vol:14} and \citet{lamb:22} applies the Lucy–Richardson (LR) algorithm independently to each of the $n$ sub-images, yielding estimates $\hat S^1,\ldots,\hat S^N$ from each pinhole. The overall sample mean source estimate, $\bar{S}$, is then obtained by pixel-wise averaging across the pinholes:
\begin{equation}\label{eq:barS}
    \bar S_{ij} = \sum_{n=1}^N \frac{\hat S^n_{ij}}{N}.
\end{equation}
The LR algorithm uses the Expectation-Maximization (EM) algorithm to estimate each pixel in the underlying image $S_{ij}$ under a Poisson observation model $Y_{lh} \sim \text{Poisson}(S_{ij}*K_{ij,lh} )$, where $S_{ij}*K_{ij,lh}$ represents the expected intensity at each pixel. The EM algorithm is an iterative procedure for maximum likelihood estimation in latent variable models that alternates between (i) the E-step: computing the expected log-likelihood given current parameter estimates, and (ii) the M-step: maximizing this expectation to update the parameters; see \citet{vol:14} for further details. The data generating and reconstruction process can be visualized in the flow chart shown in Figure \ref{fig:two_row_images}. 

In ICF applications, the true source $S$ is unobserved. Consequently, we study whether a hypothesized underlying source could have plausibly generated the observed data and whether the reconstructed source $\bar S$ preserves its shape and symmetry. To do so, one must first choose a representation of shape that is both interpretable and amenable to statistical comparison.


\subsection{Legendre Polynomial Decomposition}\label{subsec:Legpoly}

One mathematical approach to estimating object shape in images is to extract a contour representing the object's boundary, denoted $c$, and then infer the object's geometry from this contour. 
In particular, we consider \emph{percentile contours}, denoted $c_p$, defined as the ordered set of points in $\mathbb{R}^2$ corresponding to grid locations where the image intensity equals a specified percentile $g_p$ of the image’s intensity distribution $G$ (see Section~\ref{sec:method} for details).
Historically, estimation of the true source's shape in ICF image analysis is based on the $17^{\text{th}}$ percent contour of the estimated source image $\bar S$ \citep{gul:13}. Let $g_{0.17}$ denote the $17^{\text{th}}$ percentile of the values $\{\bar S_{ij} : (ij)\in\mathcal{G}^S\}$. The corresponding contour is defined as
\[
c_{0.17}
=
\left\{
X^S_{ij} \in \mathbb{R}^2 :
\bar S_{ij} = g_{0.17}
\right\}.
\] 
In ICF, this contour is typically represented via a Legendre polynomial decomposition based on a complete orthogonal polynomial basis \citep{ral:24}. The resulting basis coefficients provide a low-dimensional summary of contour geometry and help quantify symmetry.
Let $\{P_d(X^S)\}_{d=1}^D$ denote the Legendre polynomials with associated coefficients $a_d$. For a fixed $p = 0.17$, let the contour extracted from the centroid locations in the source image be denoted by $c(X^S)$ and approximated by
\begin{equation}
    c(X^S) \approx \hat c(X^S)
    =
    \sum_{d=1}^{D} a_d P_d(X^S),
\end{equation}
where the weights quantify the harmonic content of the contour. Practitioners typically understand the shape through the first five coefficients $\{a_0,\ldots,a_4\}$ where $a_0$ describes the global radius of the shape. If the shape is perfectly symmetric, or a circle, all the higher order coefficients should be zero. 

When the shape of the source is quantified in terms of this spherical harmonic fit to a selected contour, there are two main disadvantages. First, low-order Legendre polynomial representations struggle to capture localized asymmetries because the basis functions are global and smooth. High-order Legendre polynomial representations are typically numerically unstable, noise-sensitive, and difficult to interpret; they can also struggle to fit to shapes with very local artifacts (e.g., small divots). Therefore, although many shapes, such as circles and ellipsoids, fall within the scope of the spherical harmonics framework, any shape that is not smooth, simply connected, and sphere-like will not be naturally described by this representation. Table~\ref{tab:legendrecoefficents} illustrates this limitation by presenting two distinct shapes that are summarized by similar Legendre polynomial coefficients. Each row contains an example image with an associated contour (red curve) and each column contains the corresponding Legendre coefficient up to $a_5$. 


\begin{table}
\centering
\begin{tabular}{ |m{4cm}||m{1cm}|m{1cm}|m{1cm}|m{1cm}|m{1cm}|m{1cm}|}
 \hline
 \hline
Legendre Coefficients 
& $a_0$ & $a_1/a_0$ & $a_2/a_0$ & $a_3/a_0$ & $a_4/a_0$ & $a_5/a_0$ \\
 \hline
 \hline
\includegraphics[width=\linewidth]{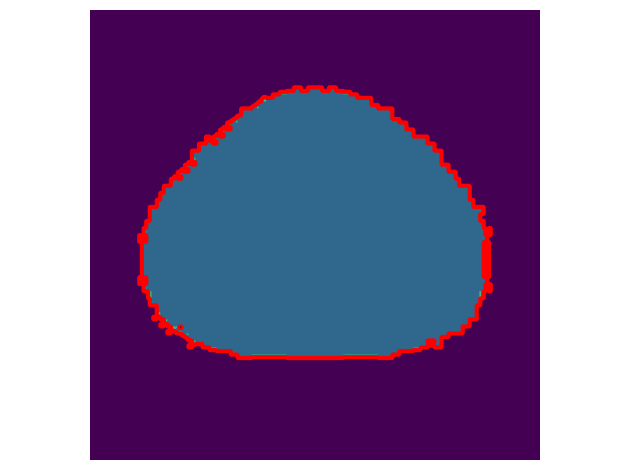} & 179 & 4\% & -17\% & -14\% & 2\% & 1\% \\
\hline
\includegraphics[width=\linewidth]{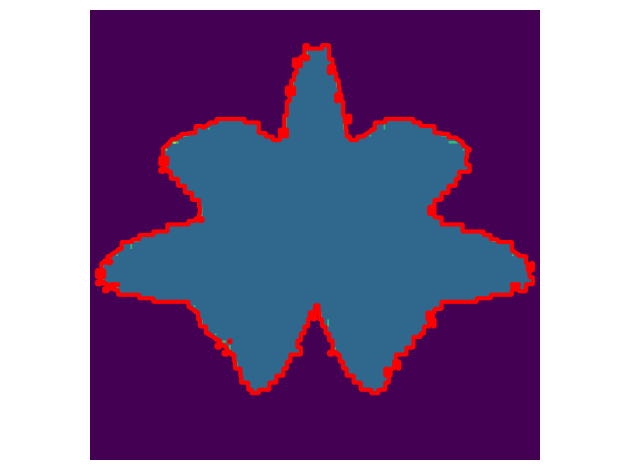}  &  179 & 5\% & -17\% & -14\% & 2\% & 1\%  \\
 \hline
\end{tabular}
\caption{Table of Legendre coefficients $a_0, a_1/a_0,\ldots,a_5/a_0$ (columns) for each example image (rows).}
\label{tab:legendrecoefficents}
\end{table}

The second main disadvantage is that interpretation and statistical inference based on the Legendre coefficients are not straightforward. Multiple coefficients may describe different aspects of a particular shape and there may be complex dependencies between these coefficients, so any inference procedure must account for this structure. 
Furthermore, it can be difficult to tie visual local shape differences to coefficient differences: e.g., a small, asymmetric divot introduced within a contour may result in a change of values across many coefficients.
These limitations motivate our use of an alternative approach from statistical shape analysis, namely ESA, which allows for more flexible methods with straightforward interpretations.

\subsection{Elastic Shape Analysis}\label{sec:esa}
In short, ESA is a statistical shape analysis approach for modeling shapes of contours. A key component underlying ESA is the definition of a metric which measures shape differences between contours. In the ICF setting, this metric can be used to compare source shape contours in $\bar S$ to a hypothesized source shape contour. 

More specifically, we assume the contours $c$ are absolutely continuous, parameterized curves that map parameter values from some input domain $\mathcal{D}$ (e.g. $[0,1]$ if open, or the unit circle $\mathbb{S}^1$ if closed) to the Euclidean plane (planar curve):
\begin{equation}\label{eq:closedcurve}
    c: \mathcal{D}  \longrightarrow \mathbb{R}^2.
\end{equation} 
The assumption of absolutely continuous contours is important for ESA, as it ensures the contour is differentiable almost everywhere. In many applications, this assumption is reasonable,
as it only excludes functions which jump or oscillate infinitely 
over small regions. 
For this work, we focus on planar closed curves, meaning the starting and ending points are the same (e.g., $c(0)=c(1)$), in alignment with our data (e.g., contours of the neutron source); however, these ideas generalize naturally to open curves, as well as curves embedded in higher-dimensional spaces \citep{kur:12, Sri:16}.

In order to encode the geometry of the curve as a smooth function, not just as a set of unordered points, $c$ can be represented as a parameterized curve $c(t) = X^S_{ij}(t) \in \mathbb{R}^2$. The parameterization dictates how the curve is traversed: the same geometric curve can be traced at different speeds, directions, and starting from different points. The $\mathbb{L}^2$  metric gives a simple, interpretable measure of overall distance between two parameterized curves $c_1, c_2$, defined by:
\begin{equation}
    \vert\vert c_1 - c_2 \vert\vert = \sqrt{\int_{\mathcal{D}} \vert c_1(t) -c_2(t) \vert^2 \ dt} \,,
\end{equation} 
where $\vert\vert \cdot \vert\vert$ is the $\mathbb{L}^2$ norm.
Unfortunately, this is not an appropriate metric for comparing \emph{shapes} of curves: two identical curves $c_1, c_2$ will have distance 0. However, if $c_2$ is re-parameterized, scaled, rotated, or translated to form a new curve $c_2^*$, its shape has not changed, but the $\mathbb{L}^2$ distance between $c_1$ and $c_2^*$ will be nonzero \citep{Sri:16}.

To account for these shape-preserving transformations, we want a curve representation along with a metric that is invariant to rigid motion (rotation and translation), global scaling, and reparameterization, only measuring intrinsic shape differences. 
One way to achieve this is by transforming curves to their {\em square root velocity function} (SRVF) representation, defined as:
\begin{equation}\label{eq:srvf}
q(t) = \frac{\dot c (t)}{\sqrt{\vert \dot c (t)\vert }},
\end{equation} 
where $\dot{c}$ is the derivative and $\vert \cdot \vert$ denotes the vector $2-$norm. \cite{Sri:11} showed that the $\mathbb{L}^2$ distance between SRVFs is equivalent to a well-studied class of \emph{elastic metrics}, which measures the amount of bending and stretching required to deform one curve into another. 

Furthermore, this naturally facilitates the definition of a metric invariant to shape-preserving transformations.  Under this representation, two curves which only differ by a translation differences are mapped to the same SRVF, allowing for translation-invariant curve comparisons. To introduce scale invariance, SRVFs can be mapped to a pre-shape space $\mathcal{C}$, which is the set of all SRVFs with unit norm:
\begin{equation}
    \mathcal{C} = \left \{ q \in \mathbb{L}^2(\mathcal{D},\mathbb{R}^2) \ \Big| \int_{\mathcal{D}} \vert  q(t) \vert dt = 1, \int_{\mathcal{D}} q(t) \vert q(t) \vert  dt = 0 \right \} \,
\end{equation}
where $\mathbb{L}^2(\mathcal{D},\mathbb{R}^2)$ is the set of all square-integrable functions from $\mathcal{D}$ to $\mathbb{R}^2$. The pre-shape space $\mathcal{C}$ forms a unit Hilbert sphere, inheriting its differential geometric structure.
Thus, SRVF comparisons on $\mathcal{C}$ are translation and scale-invariant.

To account for rotation and reparameterization invariance requires forming a quotient space, as opposed to standardization. More formally, let the reparameterization group for a closed curve be defined by $\Gamma=\{\gamma:\mathcal{D} \longrightarrow \mathcal{D} \vert \gamma(0)=\gamma(1)$, $\gamma$ is a diffeomorphism$ \}$. A reparameterization $\gamma \in \Gamma$ of contour $c$ (with SRVF $q$) yields reparameterized contour $(c \circ \gamma)(t) = c(\gamma(t))$ (with SRVF $(q \circ \gamma)\sqrt{\dot{\gamma}}$). Under the $\mathbb{L}^2$ metric, identically reparameterizing two contours by $\gamma$ changes the distance (i.e., $\vert\vert c_1-c_2 \vert\vert \neq \vert\vert c_1 \circ \gamma -c_2 \circ \gamma \vert\vert $). Crucially, however, the $\mathbb{L}^2$ distance between SRVFs is preserved: $\vert \vert q_1-q_2 \vert \vert_{\mathbb{L}^2} = \vert \vert (q_1 \circ \gamma)\sqrt{\dot{\gamma}} - (q_2 \circ \gamma)\sqrt{\dot{\gamma}} \vert \vert_{\mathbb{L}^2}$. Lastly, we define the set of all rigid rotations $SO(2)=\{O \in \mathbb{R}^{2 \times 2} \vert O^TO = I, \text{det}(O)=+1 \}$. The elastic shape space $\mathcal{S}$ is the quotient space formed by taking the pre-shape space $\mathcal{C}$ modulo rotation and reparameterization groups:
\begin{equation}
    \mathcal{S} = \mathcal{C} / \left( SO(2) \times \Gamma \right) \,.
\end{equation}  

Elements of $\mathcal{S}$ are equivalence classes which represent all forms of the elastic shape for a curve subject to shape-preserving transformations: $[q] = \{O(q \circ \gamma)\sqrt{\dot{\gamma}} \ | \ O \in SO(2), \gamma \in \Gamma, ||q||^2_{\mathbb{L}^2} = 1 \} \in \mathcal{S}$. The elastic shape distance (ESD) is a metric between two equivalence classes $[q_1],[q_2]$ on $\mathcal{S}$, defined by:
\begin{equation}\label{eq:esd}
    d_{\mathcal{S}} ([q_1],[q_2]) = \inf_{(\gamma, O) \in \Gamma \times SO(2), ||q_1||^2_{\mathbb{L}^2} = 1} \cos^{-1} \left( \Big\langle\Big\langle q_1,O(q_2 \circ \gamma \sqrt{\dot \gamma} \Big\rangle\Big\rangle_{\mathbb{L}^2} \right) \,,
\end{equation}
where $\langle\langle \cdot,\cdot \rangle\rangle_{\mathbb{L}^2}$ is the $\mathbb{L}^2$ inner product.
\cite{Sri:11} proved that ESD is a valid metric on $\mathcal{S}$, inheriting its form from the differential geometric structure of $\mathcal{C}$.
Central to this definition is the role of \emph{registration}: finding the optimal alignment (reparameterization and rotation) of SRVF $q_2$ to best match reference SRVF $q_1$ under the elastic metric. There is no closed-form solution to this; thus, numerical optimization is used by alternating between finding the optimal rotation (fixing reparameterization) and optimal reparameterization (fixing rotation) until a stable minima ($O^*,\gamma^*$) has been reached. Algorithmic details can be found in \cite{Sri:16}.


\begin{figure}
\centering
    \centering
    \includegraphics[width=\linewidth]{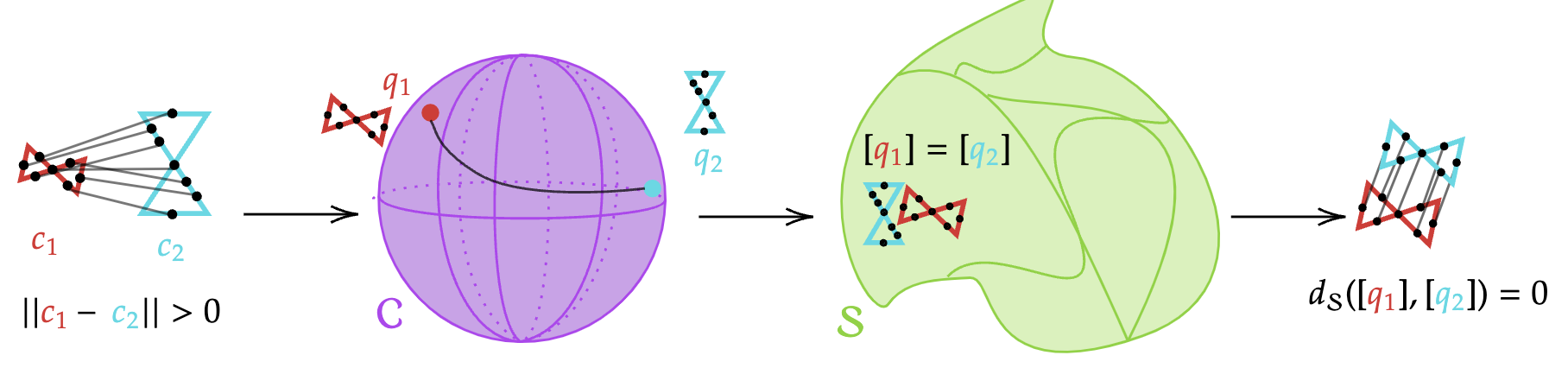}
\caption{Example of pipeline to get ESD between two curves ${c_1},{c_2}$ with the same shape but different rotation, scale, parameterization, and translation.} \label{fig:ESA_intuition}
\end{figure} 

Figure~\ref{fig:ESA_intuition} visually describes ESD between two separate curves ${c_1},{c_2}$ with the same shape but different scales, rotations, locations, and parameterizations where $\vert\vert {c_1} -{c_2}\vert\vert \geq 0$. The curves are first mapped to the pre-shape space ${\mathcal{C}}$ where differences due to scale and location are removed with the SRVF representations ${q_1}, {q_2}$. Then under the equivalence relation of rotation and parameterization, the two SRVFs lie in the same equivalence class $[{q_1}],[{q_2}]$ in the shape space ${\mathcal{S}}$ so that ${q_1}(t) = O^*({q_2}\circ \gamma^*)(t)\sqrt{\dot \gamma^*(t)} $ for some $O^*$ and $\gamma^*$. The ESD between $c_1$ and $c_2$ is $d_{\mathcal{S}}([q_1],[q_2]) = 0$ since they both have the same intrinsic shape.

Although the ESA framework provides a flexible and computationally efficient way to quantify shape differences via the ESD, its use for inference on a contour extracted from image data (e.g., for asking whether some source could have plausibly generated an observed image contour) requires additional care. In particular, the ESD is defined between two curves, but in our setting the key challenge is inferential: we typically observe only a single estimated contour (e.g., from the sample mean image) and must characterize the distribution of its ESD relative to a hypothesized source contour. Moreover, beyond tangent-space approximations, there are no explicit distributional assumptions for the raw ESDs, which limits parametric inference. Even common nonparametric approaches in shape analysis (e.g., \cite{fas:14, ama:10}), such as the empirical bootstrap, can fail when the overall mapping from the image data to the ESD (via contour extraction and alignment; see Equation~\eqref{eq:esd}) is not continuously differentiable.

\section{Bootstrap Inference of ESD Distributions}\label{sec:method}

In this section, we develop 
a bootstrap procedure to estimate uncertainty in the ESD, called {\bf bootESA}. This section is divided into three parts: a general overview of the methodology (Section~\ref{subsec:bootESAmethod}), the mathematical justification (Section~\ref{subsec:mathmaticalFramework}), and its application to ICF data in particular (Section~\ref{subsec:ICFmethod}). The notation used for Section~\ref{subsec:bootESAmethod} and Section~\ref{subsec:mathmaticalFramework} is summarized in Table~\ref{tab:notation} to demonstrate the generalizability of the method.

\begin{table}[ht]
\centering
\caption{General notation and corresponding examples for the data application.}
\label{tab:notation}
\begin{tabular}{lll}
\hline
\hline
Symbol & Meaning & ICF data example \\
\hline
\hline
$Y^1,\ldots,Y^N$ & Data & Pinhole images \\
$S$ & Underlying data space & Source image \\
$\bar S$ & Learned mean space & Sample mean image \\
$\bar S^*_b$ & Bootstrap mean spaces & Bootstrap sample mean images \\
$c \text{ (or } q)$ & Curve (or SRVF) of object in $S$ & Source contour (or SRVF) \\
$\bar c  \text{ (or } \bar q)$ & data curve (or SRVF) of object & Sample mean contour (or SRVF) \\
$\bar c^*_b \text{ (or } \bar q^*_b)$ & Bootstrap curves (or SRVFs) of object & Bootstrap contours (or SRVFs) \\
\hline
$c^0 \text{ (or } q^0)$ & Hypothesized shape (or SRVF) & Circular source contour (or SRVF) \\
\hline
\hline
\end{tabular}\label{tab:notation}
\end{table}


\subsection{bootESA Method}\label{subsec:bootESAmethod}

The {\bf bootESA} method can be broken down into four steps for any general data set $\{Y^1,\ldots,Y^N \}$ that generates some learned mean $\bar S$ of a true space of interest $S$. This procedure allows us to make an inferential statement about the shape of objects in $S$ for a single prespecified percentile contour.

\subsubsection{Bootstrap}\label{subsub:step1}

In this step, a bootstrap procedure generates pseudo samples of the learned mean $\bar S$. We sample $m \leq N$ data observations with replacement $\{ Y^{1*},\ldots,Y^{m*} \}_b$ where $b\in\{1,\ldots,B\}$ indexes bootstrap samples  and the corresponding bootstrapped mean $\bar S^*_b$. 
The standard nonparametric bootstrap uses resamples of size $m=N$.
However, there may be instances where selecting $m<N$ is necessary;
we address how to select $m$ in Section~\ref{subsubsec:moutofn}. 
\\
{\bf input:} $\{Y^1,\ldots,Y^N \}$\\
{\bf output:} $\{ \bar S^*_1, \ldots, \bar S^*_B \}$
 
\subsubsection{Compute ESD}\label{subsub:step2}

In this step, we characterize shape uncertainty due to sampling variability by constructing a distribution of ESDs, obtained by extracting shapes of interest from mean and bootstrapped mean images and then measuring their distance using the elastic metric.



Recall that the image is on some grid $\mathcal{G}$ with $i\in\{1,\ldots,n_i\}$ columns and $j\in\{1,\ldots,n_j\}$ rows such that each grid cell $(ij) \in \mathcal{G}$ can be represented by its centroid (i.e.,  $X_{ij} \in \mathbb{R}^2$). We first map the image to a planar curve, given by a percentile contour at fixed percentage $p$: let $f_p: \mathbb{R}^{n_{i} \times n_{j}} \longrightarrow \mathbb{R}^2$ represent this function.
Specifically, the $p^{\text{th}}$ percentile of the distribution of pixel intensities in mean image ($\bar S_{ij} \overset{iid}{\sim} G$) is given by $g_p = G^{-1}(p)$, and the percentile contour is an unordered set of $(x,y)$ coordinates where the intensity value is equal to the percentile: \[ f_p(\bar S) \equiv \bar c_p = \{X_{ij} : \bar S_{ij}=g_p, (ij) \in \mathcal{G} \} \,. \]

Then, we represent the contour by its SRVF, mapping it to the pre-shape space; let $f_{\mathcal{C}}: \mathbb{R}^2 \longrightarrow \mathcal{C}$ be this function 
defined by Equation~\eqref{eq:srvf}.
Finally, we compute the ESD between this SRVF and a reference SRVF $q_p$ (e.g., a hypothesized target shape). In particular, let $f_{q_p}: \mathcal{C} \longrightarrow \mathbb{R}$ be a function which maps a SRVF to the shape space $\mathcal{S}$ and then calculates the ESD (Equation~\eqref{eq:esd}) with reference SRVF $q_p$. In other words: $f_{q_p}(\cdot) = d_{\mathcal{S}}([q_p],[\cdot])$.

Let $f(S|q_p) = f_{q_p}(f_{\mathcal{C}}(f_{p}(S))$ be the full function mapping the image to the ESD. Using the SRVF of the mean image $\bar S$ as a stand-in for the SRVF of the true space $S$, and the bootstrapped mean images $\bar S^*$ as stand-ins for new realizations of the data $\bar S^*$, we generate a distribution of distances: $\{f(\bar S^*_{1}|\bar q_p), \ldots, f(\bar S^*_{B}|\bar q_p) \}$.
\\
{\bf input:} $\bar S, \{ \bar S^*_1, \ldots, \bar S^*_B \}$ \\
{\bf output:} $\{d_{\mathcal{S}}([\bar q_p],[\bar q^*_{p,1}]), \ldots, d_{\mathcal{S}}([\bar q_p],[\bar q^*_{p,B}])\}$

\subsubsection{Confidence Interval}\label{subsub:step3}
The third step of the {\bf bootESA} method creates a confidence interval from the empirical bootstrap resamples of the ESD after an uncertainty-calibrating rescaling (i.e., based on knowing or estimating $m$). 
The true confidence interval for the ESD between the contour of the object in the underlying space and contours of estimated space is defined as follows:
\begin{equation}\label{eq:ci_true}
    C_{\alpha, m} = \{[q_p] : \sqrt{m} \cdot d_{\mathcal{S}}([q_p],[\bar q_p]) < Z_{\alpha} \},
\end{equation}
where $Z_{\alpha}$ is the $1-\alpha$ percentile of the ESD distribution and $\sqrt{m}$ is the known asymptotic scale. We approximate the confidence interval with $\hat C_{\alpha, m}$ which replaces $Z_\alpha$ with $\hat Z_{\alpha}$ via bootstrap resamples:
\begin{equation} \label{eq:CI}
    \hat Z_{\alpha} = \inf \left \{z: \frac{1}{B} \sum_{b=1}^{B} I((R_m^*)_b > z) < \alpha \right \}
\end{equation}
where $(R_m^*)_b = \sqrt{m} \cdot d_{\mathcal{S}}([\bar q_{p}], [\bar q^*_{p,b}])$.
\\
{\bf input:} $\{d_{\mathcal{S}}([\bar q_p],[\bar q^*_{p,1}]), \ldots, d_{\mathcal{S}}([\bar q_p],[\bar q^*_{p,B}])\}$ \\
{\bf output:} $\hat C_{\alpha,m} = [0, \hat Z_{\alpha}]$

\subsubsection{Hypothesis test}\label{subsub:step4}
The confidence interval described in Equation~\ref{eq:CI} provides a natural hypothesis testing framework by rejecting the null hypothesis whenever the hypothesized parameter value lies outside the region. We start with a hypothesized shape for the object of interest denoted by $c^0$ and compute its SRVF $q^0$. The null and alternative hypotheses are stated below:
\begin{equation}\label{eq:null}
\begin{split}
    H_0 : [q^0] \in  C_{\alpha,m}\\
    H_A : [q^0] \notin C_{\alpha,m}
\end{split}
\end{equation}

The test statistic is defined as $R_m^{\text{obs}} = \sqrt{m} \cdot d_{\mathcal{S}} ([q^0],[\bar q_p])$ and our bootstrap distribution is $(R_{m}^*)_b = \sqrt{m} \cdot d_{\mathcal{S}} ([\bar q_p],[\bar q_{p,b}^*])$ where $b=\{1,\ldots, B\}$ indexes the $B$ bootstrap samples. The $p$-value is calculated using the empirical distribution of distances $\hat F^*=R_{m}^*$ and counting the proportion of $R_{m}^*$ values that are larger than our test statistic.

\begin{equation}\label{eq:pval}
    p\text{-value} = \frac{\sum_{b=1}^{B} \mathbb{I} ((R_{m}^*)_b > R_m^{\text{obs}})}{B}.
\end{equation}
\\
{\bf input:} $c^0, \hat C_{\alpha,m} = [0, \hat Z_{\alpha}]$ \\
{\bf output:} $p$-value with decision (either Reject or Fail to Reject).

\subsection{Mathematical Framework bootESA}\label{subsec:mathmaticalFramework}

In this section, we justify the {\bf bootESA} method and describe how to find the correct $m$ and asymptotic scale for the limiting distribution of ESDs. From step one, given a dataset $\{Y^1,\ldots,Y^N\}$ of sample size $N$, we can get an estimate of the true space $S$ with a learned mean $\bar S$. The estimator of each pixel is a sample mean, so under standard regularity conditions and by the law of large numbers, each pixel is a consistent estimator of the corresponding pixel in the true image. Therefore $\forall X_{ij} \in \mathcal{G}$,
\begin{equation}\label{eq:pixelconvergence}
    \bar S_{ij} \overset{prob}{\longrightarrow}  S_{ij}.
\end{equation}
The sample mean image $\bar S$ is a consistent estimate of the true image $S$ assuming a fixed, finite number of pixels so that
\begin{equation}\label{eq:convergenceImages}
    \bar S \overset{prob}{\longrightarrow}  S.
\end{equation}
Under the Multidimensional Central Limit Theorem, the limiting distribution of the sample mean image is
\begin{equation}\label{eq:joint}
    \sqrt{N} \left(
\begin{bmatrix}
\bar S_{1,1} \\
\vdots   \\
\bar S_{n_{i},n_{j}} 
\end{bmatrix}
 - \begin{bmatrix}
S_{1,1}  \\
\vdots   \\
 S_{n_{i},n_{j}}
\end{bmatrix}\right) \overset{prob}{\longrightarrow}  \mathcal{MVN}(0,\Sigma),
\end{equation}
where $\Sigma = \sigma^2_{ij,i'j'}$ is the covariance matrix for any $\bar S_{ij}$ and $\bar S_{i'j'}$. Therefore, jointly, $\bar S - S = O_p \left(\frac{1}{\sqrt{N}} \right)$ in $\mathbb{R}^{n_{i}\times n_{j}}$.

The bootstrap method outlined in Section~\ref{subsub:step1} samples the data with replacement to generate bootstrap samples $Y^{1*}, \ldots, Y^{m*}$ where $m=N$ by Equation~\ref{eq:joint}. The bootstrap sample means are denoted by $\bar S^*_b$ for $b=\{1,\ldots,B \}$. This allows for inference on the entire image with a valid bootstrap procedure meaning,
\begin{equation}
    \sqrt{N}(\bar S^* - \bar S) \overset{d}{\approx} \sqrt{N}(\bar S - S)).
\end{equation}

However, the goal of the {\bf bootESA} method is to infer the distribution of the ESD between contours in images as opposed to performing inference on the image itself. Accordingly, the {\bf bootESA} method must approximate the following limiting distribution:
\begin{equation}
    \sqrt{m}(f(\bar S| q_p) - f(S| q_p)) = \sqrt{m}(d_{\mathcal{S}}([q_p],[\bar q_p]) - d_{\mathcal{S}}([q_p], [q_p]))=\sqrt{m}(d_{\mathcal{S}}([q_p],[\bar q_p])),
\end{equation}
via bootstrap resampling. Using the bootstrap resamples with $m=N$ and asymptotic scale of $\sqrt{N}$ in the confidence region (Section~\ref{subsub:step3}) and hypothesis test (Section~\ref{subsub:step4}) is valid when $f(\cdot|q_p)$ is a continuously differentiable function in a neighborhood of $S$. With a continuous differentiable function, we know that $f(\bar S|q_p) \overset{prob}{\longrightarrow} f(S|q_p)$ by the continuous mapping theorem (CMT) and $f(\bar S|q_p) - f(S|q_p) = O_p\left( \frac{1}{\sqrt{N}} \right)$ by the Delta Method. 

Continuous differentiability is a vital assumption to know if we have stability (i.e., small changes in the data lead to small changes in the distribution of the ESD). Under strict data assumptions, the function mapping the image to the SRVF (i.e., $f_{\mathcal{C}}(f_p(S))$) is continuously differentiable whereas $f_{q_p}(q)$ is directionally differentiable (i.e., undefined derivative at the origin for ESD). We discuss the assumptions needed for the SRVFs to be continuously differentiable before discussing the directional differentiability of the ESD between a curve and itself.

\begin{assumption}\label{as:smooth}
    The distribution of pixels in the sample mean image, denoted by $\bar S_{ij} \overset{iid}{\sim} G$, must be smooth and non-degenerate.
\end{assumption}

\begin{assumption}\label{as:topology}
    The topology of the contour is assumed to be fixed at one closed curve and this does not change for small perturbations in the data.
\end{assumption}

The smoother the image, the more likely both of these assumptions are to hold. A smooth image induces a smooth distribution over pixels, implying that the percentile function $g_p$ is differentiable, and it also results in smoother contours. Moreover, sufficient smoothness of the planar curve guarantees that the SRVF is continuously differentiable \citep{Sri:16}. The other constraint to ensure that $f_{p}(S)$ is continuously differentiable is that the percentage $p$ cannot be on the boundary of the distribution of pixels $G$ (e.g $p \neq 0,1$).

The second assumption is needed because we can only compute the ESD between two curves, meaning the topology cannot change. This assumption generally holds for smooth images, large sample sizes, and certain percentile levels. In practice, when more than one curve is found in an image at a given percentile level, we added a step to select the contour with the largest area.


As long as Assumptions~\ref{as:smooth} and \ref{as:topology} hold, the part of the function $f$ which maps an image to the pre-shape space $\mathcal{C}$ is continuously differentiable at $S$. By the CMT we can say the unordered set of points making up the contour and SRVFs are consistent estimates of the true contour and SRVF: $\bar c_p \overset{prob}{\longrightarrow} c_p$ and $\bar q_p \overset{prob}{\longrightarrow} q_p$.

Therefore, based on Equation~\eqref{eq:joint} and continuous differentiability of $f_{\mathcal{C}}(f_p(S)) = q_p$ the random perturbation of size $N^{-1/2}$ propagates through $f_{\mathcal{C}}(f_p(S))$ so that $\bar q - q = O_p \left(\frac{1}{\sqrt{N}}\right)$ assuming that the contours have the same order, direction, and starting points. By the Delta Method with a nonzero Jacobian $f_{\mathcal{C}}'(c_p)$:
\begin{equation}\label{eq:SRVFconvergence}
    \sqrt{N} (f_{\mathcal{C}}(\bar c_p) - f_{\mathcal{C}}(c_p)) \longrightarrow f_{\mathcal{C}}'(c_p)Z,
\end{equation}
and the bootstrap procedure is valid so that:
\begin{equation}
    \sqrt{N}(\bar q^*_p - \bar q_p) \overset{d}{\approx} \sqrt{N}(\bar q_p - q_p).
\end{equation}

However, we do not have continuous differentiability when moving from the SRVF to the shape space $\mathcal{S}$. The first issue is the optimization step (e.g., $\inf_{(\gamma,O) \in \Gamma \times SO(2)}$ in Equation~\eqref{eq:esd}) where there is no guarantee of a unique optimal solution $(\gamma^*,O^*)$. Even assuming that some unique solution exists for the optimization step $(\gamma^*,O^*)$ such that $q=q_1=q_2$, the derivative $\frac{d}{dq} \cos^{-1} \Big( \Big\langle\Big\langle q,q \Big\rangle\Big\rangle_{\mathbb{L}^2} \Big)$ is undefined.

In other words, as $N \longrightarrow \infty$, the data SRVF $\bar q_p$ and bootstrap SRVFs $\bar q_p^*$ converge to the true SRVF $q_p$. However, the bootstrap distribution of ESD (which is moving to zero) can exhibit inflated variability and be less concentrated near zero. This motivates use of a bootstrap procedure called the $m$-out-of-$N$ bootstrap method.

\subsubsection{$m$-out-of-$N$ Bootstrap Procedure}\label{subsubsec:moutofn}
One approach to bootstrapping non-smooth functions (e.g. when the functional delta method fails) is using the $m$-out-of-$N$ bootstrap procedure \citep{bic:08, bic:11}. In this procedure, $m$ observations are sampled with replacement from the data set of size $N$ where $m = N^{\epsilon} $ and $\epsilon \in (0,1)$. The only assumption for this setting is that $m/N \longrightarrow 0$ as $m \longrightarrow \infty$ and the closer the function is to being differentiable (i.e. the closer the bootstrap is to being valid) the higher the $\epsilon$ value should be. The choice of $m$ has historically been guided by extrapolation-based arguments. More recent practice often considers a range of $m$ values, increasing $m$ until a chosen distance between successive distributions becomes small and remains stable \citep{bic:08, pol:11}. We discuss this procedure further in Section~\ref{subsec:sim1} and in the supplemental material.

The only adjustments needed to the {\bf bootESA} method outlined in Section~\ref{subsec:bootESAmethod} are (i) the test statistic uses $\sqrt{N}$ so that $R_N^{\text{obs}} = \sqrt{N} \cdot d_{\mathcal{S}}([q], [\bar q])$ and (ii) to estimate the correct $\epsilon$ for $\sqrt{m}=\sqrt{N^{\epsilon}}$ so the bootstrap distribution is $R_m^* = \sqrt{m} \cdot d_{\mathcal{S}}([\bar q],[\bar q^*])$.

\subsection{ICF data bootESA}\label{subsec:ICFmethod}

Recall from Section~\ref{subsec:datasetup} that the estimate $\bar S$ of the source uses realized images from $N$ pinholes: $\{Y^1,\ldots, Y^N \}$ \citep{vol:14, lamb:22}. Conditional on known point spread functions  $\{K^n\}_{n=1}^N$ for all pinholes, the reconstructed source images $\hat S^1,\ldots,\hat S^N$ are independent and identically distributed draws from some distribution $F$ with mean image equal to the true source $S$; that is, $\mathbb{E}_F[\hat S^n|K^n]=S$ for all $n\in\{1,\ldots,N\}$.

Traditional ICF data images can be assumed to be $iid$ since the relative sizes of the pinholes in the aperture are much larger than the true source (i.e., the pinholes all view the entire source). This means that even though each pinhole forms its own image on a different detector region and photon arrivals, the correlations across pinholes are negligible. We sample pinhole images with replacement to generate bootstrap samples $\{\hat S^{*1}, \ldots, \hat S^{*m}\}_b$, from which we compute the bootstrap sample mean $\bar S^*_b$ for $b=\{1,\ldots,B\}$. 

Under regularity conditions and the EM algorithm, each pixel in the sample mean image $\bar S_{ij}$ (Equation~\eqref{eq:barS}) is a consistent estimate of corresponding pixel in the true source image \citep{wal:49} so the sample mean image $\bar S$ is a consistent estimate of the true source $S$.

\begin{figure}
\centering
    \centering
    \includegraphics[width=\linewidth]{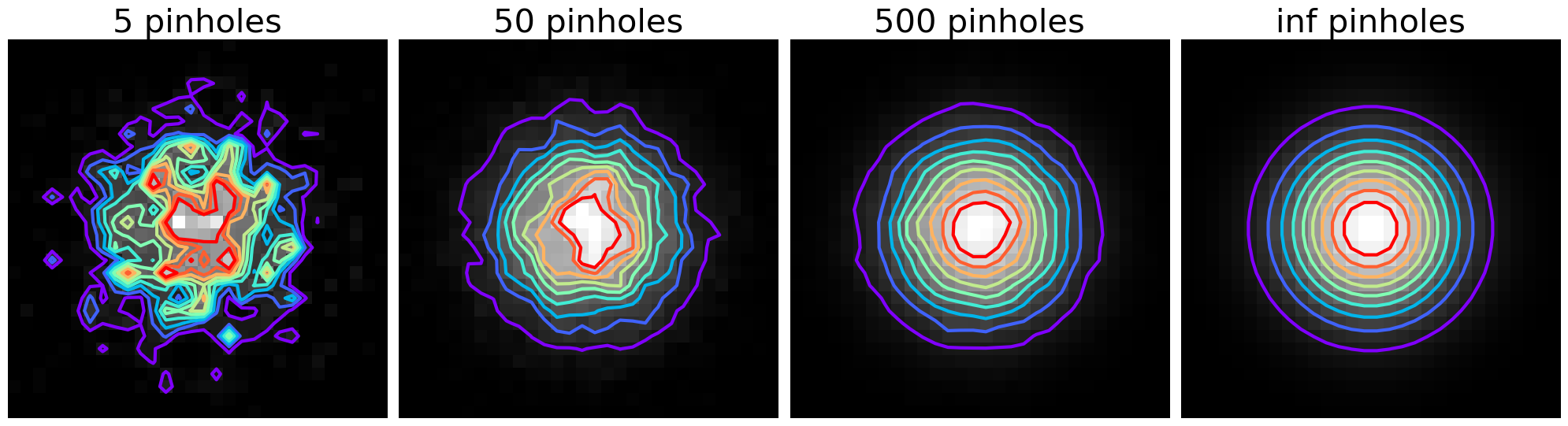}
\caption{Simulation example with images of the mean reconstructed source by number of pinholes ($N=\{5,10,50,500,\infty\}$) where the last image is the true source the data images were generated from. The different contour colors correspond to different percentile levels $p=\{0.1, 0.2, \ldots, 0.9\}$.} \label{fig:ExamplesLoops}
\end{figure}

Assumption~\ref{as:smooth} is reasonable, as ICF images tend to be generally smooth as seen in Figure~\ref{fig:realdata}. Assumption~\ref{as:topology} only fails in noisy simulations with extremely small sample sizes as seen in Figure~\ref{fig:ExamplesLoops}. This figure displays example sample means of reconstructed images for different numbers of pinholes. The data is all generated from the same underlying source seen in the last image (e.g. $N=\infty$). On top of each image are the different percentile contours where each color corresponds to percentile level $p$. Only sample sizes $N \lesssim 30$ have contours with a different topology than one connected components as seen in $N=5$.

\section{Simulations}\label{sec:simulations}

This section presents two synthetic experiments that demonstrate the accuracy, precision, and power of the hypothesis testing procedure outlined in Section~\ref{sec:method}. Section~\ref{subsec:sim1} looks at convergence of the bootstrap ESD and the true empirical ESD between the source shape and new realizations of sample mean images as a proxy for assessing the behavior of the Type I error rate $\alpha$. If the traditional bootstrap procedure is valid, as the sample size $N$ (e.g. number of pinhole sub-images) increases the distribution of distance should both converge to $0$ in line with Equation~\eqref{eq:SRVFconvergence}. However, as discussed, we use the $m$-out-of-$n$ bootstrap to account for instability resulting from the ESD operator's lack-of-smoothness; thus, we explore how varying $m$ impacts performance. Section~\ref{subsec:sim2} evaluates the test’s ability to detect departures from the null hypothesis.

For both simulation studies, we assume the true source image contains a circle as seen in Figure~\ref{fig:ExamplesLoops} for $N=\infty$, and use this to construct $100,000$ different data images $Y^1, \ldots, Y^{100,000}$ under the same data-generating process described in Section~\ref{subsec:datasetup}. By applying the EM algorithm, we obtain $100,000$ reconstructed sources $\hat S^1,\ldots, \hat S^{100,000}$. For each sample size $N$ we generate $50$ sample mean estimate: $\{\bar S^{1},\ldots, \bar S^{50}\}$ and using the steps of the {\bf bootESA} method get corresponding contour $\{\bar c^1_p,\ldots, \bar c^{50}_p\}$ and SRVF $\{\bar q^1_p,\ldots, \bar q^{50}_p\}$. The sample mean estimate used in constructing the approximate bootstrap distribution is denoted by $\bar S_{\boldsymbol{.}}$, which we take to be $\bar S_{\boldsymbol{.}} = \bar S^{1}$ for simplicity. The $B$ bootstrap images are samples from the reconstructed images $\{\bar S^1, \ldots, \hat S^N \}$ which are used to compute $\bar S_{\boldsymbol{.}}$, resulting in bootstrap sample mean images: $\{(\bar S_{\boldsymbol{.}}^*)_1,\ldots,(\bar S_{\boldsymbol{.}}^*)_B\}$. The percentile contours for the sample mean estimate and the bootstrap sample means are denoted by $\bar c_{\boldsymbol{.},p}$ and $\{(\bar c_{\boldsymbol{.},p}^*)_1,\ldots, (\bar c_{\boldsymbol{.},p}^*)_B \}$ with corresponding SRVFs $\bar q_{\boldsymbol{.},p}$ and $\{(\bar q_{\boldsymbol{.},p}^*)_1,\ldots, (\bar q_{\boldsymbol{.},p}^*)_B \}$, respectively. Therefore, the true empirical ESDs are denoted by $d_{\mathcal{S}}([q_{p}],[\bar q_{p}])$ and the approximate ESDs are denoted by $d_{\mathcal{S}}([\bar q_{\boldsymbol{.},p}],[\bar q^*_{\boldsymbol{.},p}])$. For visualization purposes, we set $p=0.65$ though multiple percentiles are considered for the Type I error assessment.


\subsection{Simulation: Asymptotic Convergence}\label{subsec:sim1}

\begin{figure}
\centering
    \begin{subfigure}{0.32\linewidth}
    \centering
    \includegraphics[width=\linewidth]{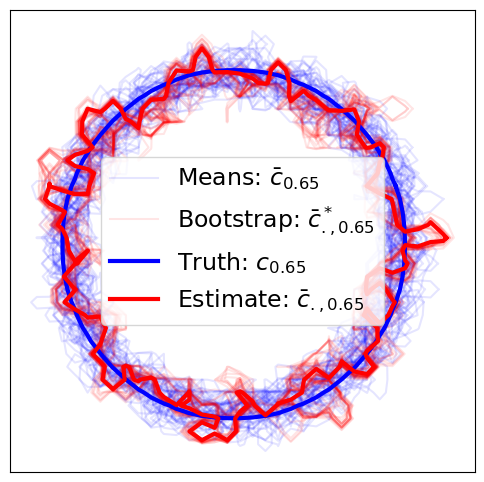}
    \caption{contours for $N=10$}\label{subfig:meancountours1}
    \end{subfigure}
    \begin{subfigure}{0.32\linewidth}
    \centering
    \includegraphics[width=\linewidth]{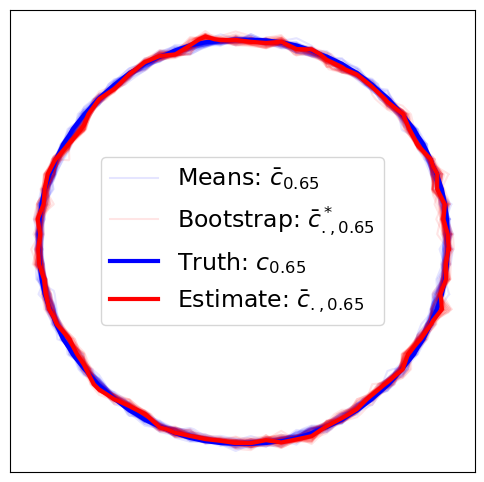}
    \caption{contours for $N=1000$}\label{subfig:meancountours2}
    \end{subfigure}
    \begin{subfigure}{0.3\linewidth}
    \centering
    \includegraphics[width=\linewidth]{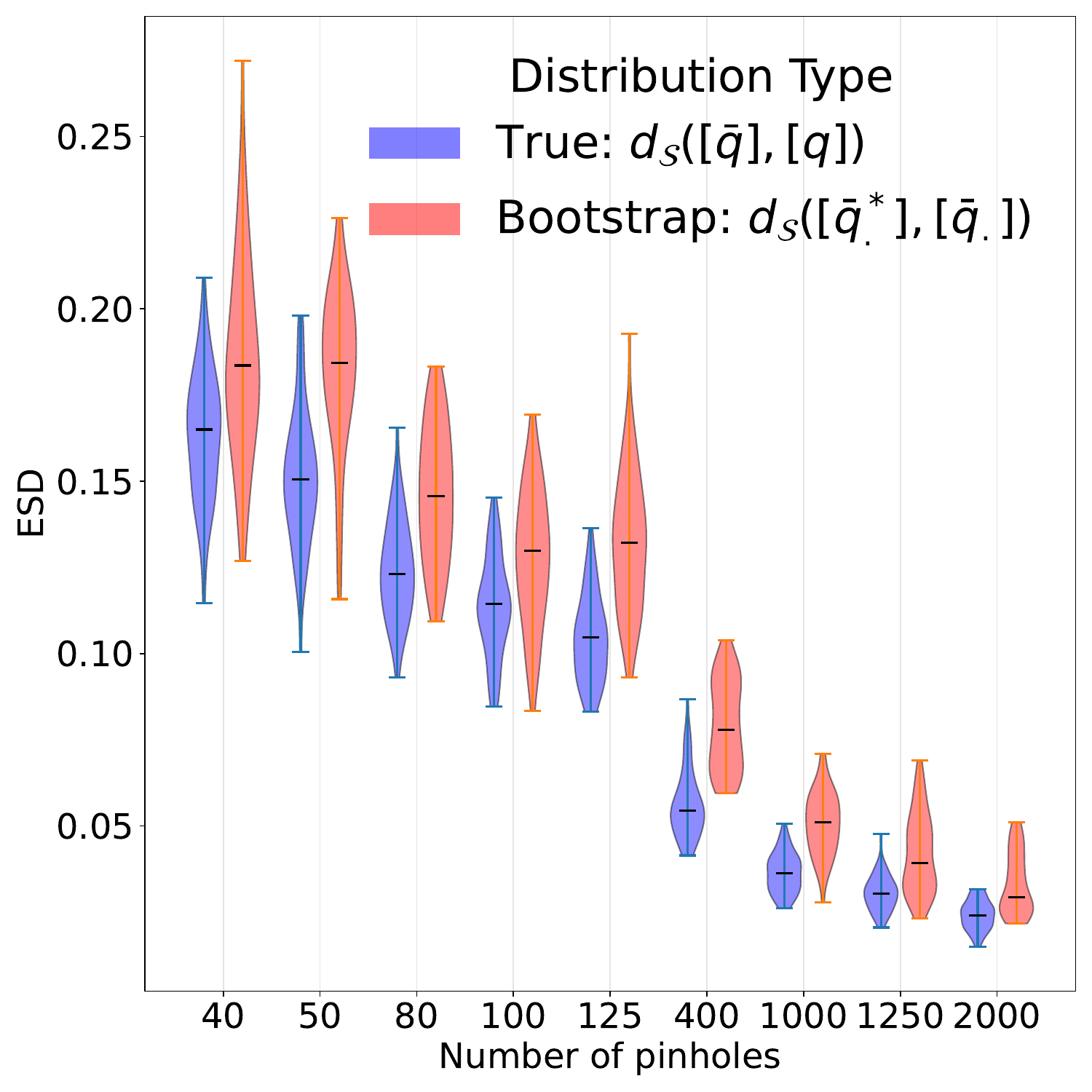}
    \caption{Violin plots ESD}\label{subfig:distances}
    \end{subfigure}
\caption{Simulation results for sample mean contours $\bar c_{0.65}$ and the source contour $c_{0.65}$ are in blue; the contour of the mean $\bar c_{\boldsymbol{.},0.65}$ and bootstrap contours $\bar c_{\boldsymbol{.}, 0.65}^*$ are in red with sample sizes of (a) $10$ pinholes and (b) $1000$ pinholes. (c) The distribution of raw ESDs for the true empirical distribution (blue) and the approximate empirical distribution (red) for sample size (x-axis).} \label{fig:simulationresults}
\end{figure}

In the first simulation study, the limiting distribution of the ESD is assessed across six different sample sizes $N=\{10,50,100,\allowbreak 500,1000,2000\}$ where the limiting distribution is comprised of $50$ observations. 
Examples of contours of the sample mean images $\bar c_{0.65}$ (light blue) and bootstrapped mean images $\bar c_{\boldsymbol{.},0.65}^*$ (light red) for the different sample sizes can be seen in Figure~\ref{subfig:meancountours1} for $N=10$ pinholes and Figure~\ref{subfig:meancountours2} for $N=1000$ pinholes. The contour of the true source, $c_{0.65}$ is blue, and the contour of the sample mean used to generate the bootstrapped images is red, $\bar c_{\boldsymbol{.},0.65}$. As seen in the figure, the contours vary less substantially as the number of pinholes increases, and the estimated contour visually appears more similar to the true source contour.

\begin{table}[h!]
\centering
\begin{tabular}{ |p{2.3cm}|p{2.3cm}||p{2cm}|p{2cm}|p{2cm}|p{2cm}| }
\hline
    \multicolumn{6}{|c|}{Type I Error Analysis $\alpha=0.1$ for $d_{\mathcal{S}} ([\bar q_p], [\bar q_p^*])$ for $\sqrt{N}$ and $\sqrt{N^\epsilon}$ for $\epsilon \in [0.6,0.9]$} \\
\hline
\bf{percentiles}
& \bf{sample size}
& $N=50$
& $N=100$
& $N=500$
& $N=1000$ \\
\hline
\hline

\multirow{5}{*}{$p=0.65$}
 & $\sqrt{N^{0.6}}$  & 0.72 & 0.74 & 0.34 & 0.32\\
\cline{2-6}
 & $\sqrt{N^{0.7}}$ & 0.38 &  0.30  &  {\bf 0.10}  & 0.16 \\
\cline{2-6}
 & $\sqrt{N^{0.8}}$  & {\bf 0.08}  & {\bf 0.06}  & {\bf 0.06}  & {\bf 0.00} \\
\cline{2-6}
 & $\sqrt{N^{0.9}}$  & {\bf 0.02}  & {\bf 0.00} &{\bf 0.00} & {\bf 0.00} \\
\cline{2-6}
 &  $\sqrt{N}$ & {\bf 0.00} & {\bf 0.00} & {\bf 0.00} & {\bf 0.00} \\
\hline

\multirow{5}{*}{$p=0.75$}
 & $\sqrt{N^{0.6}}$ & 0.62 & 0.66 & 0.20  & 0.14 \\
\cline{2-6}
 & $\sqrt{N^{0.7}}$ & 0.36 & 0.38  & {\bf 0.08} & {\bf 0.04} \\
\cline{2-6}
 & $\sqrt{N^{0.8}}$ & {\bf 0.04} & 0.12  & {\bf 0.00} & {\bf 0.02} \\
\cline{2-6}
 & $\sqrt{N^{0.9}}$ & {\bf 0.00} & {\bf 0.02} & {\bf 0.00} & {\bf 0.00} \\
\cline{2-6}
 & $\sqrt{N}$ & {\bf 0.00} & {\bf 0.00} & {\bf 0.00} & {\bf 0.00} \\
\hline

\multirow{5}{*}{$p=0.85$}
 & $\sqrt{N^{0.6}}$ & 0.62 & 0.64 & 0.28 & {\bf 0.10 } \\
\cline{2-6}
 & $\sqrt{N^{0.7}}$ & 0.34  & 0.44  & {\bf 0.08} & {\bf 0.02}   \\
\cline{2-6}
 & $\sqrt{N^{0.8}}$ & 0.16  & 0.22  & {\bf 0.00} & {\bf 0.00} \\
\cline{2-6}
 & $\sqrt{N^{0.9}}$ & {\bf 0.02} & {\bf 0.06}  & {\bf 0.00} & {\bf 0.02} \\
\cline{2-6}
 & $\sqrt{N}$ & {\bf 0.00} & {\bf 0.00} & {\bf 0.00} & {\bf 0.00} \\
\hline

\multirow{5}{*}{$p=0.95$}
 & $\sqrt{N^{0.6}}$ & 0.88 & 0.50 & 0.40 & 0.32 \\
\cline{2-6}
 & $\sqrt{N^{0.7}}$ & 0.70 & 0.50 & 0.42  &  0.28 \\
\cline{2-6}
 & $\sqrt{N^{0.8}}$ & 0.44 & 0.42 & 0.30   & 0.18 \\
\cline{2-6}
 & $\sqrt{N^{0.9}}$ & 0.32 & 0.26  & 0.20 & {\bf 0.02} \\
\cline{2-6}
 & $\sqrt{N}$ & {\bf 0.08} & {\bf 0.00} & {\bf 0.02} & {\bf 0.02} \\
\hline

\end{tabular}
\caption{Type I Error for different sample size ($N$), scales ($\sqrt{m}=\sqrt{N^\epsilon}$), and percentile ($p$) for 50 simulation runs with true source contour of a circle as seen in Figure~\ref{fig:ExamplesLoops}. We bold the values where $\hat \alpha  < \alpha=0.1$.}
\label{tab:type1error}
\end{table}
The violin plots in Figure~\ref{subfig:distances} show the ESD distributions for each sample size where the blue is the true distribution of ESD $d_{\mathcal{S}}([q_{0.65}], [\bar q_{0.65}])$ and the red is the bootstrap approximation $d_{\mathcal{S}}([\bar q_{\boldsymbol{.},0.65}], [\bar q_{\boldsymbol{.},0.65}^*])$. In this setting, the bootstrap distribution appears to upper bound the true source distribution, which can result in overly-conservative tests through decreased Type I error $\leq \alpha$ and power. 

Table~\ref{tab:type1error} presents a Type I Error analysis using the hypothesis testing procedure outlined in Section~\ref{subsub:step4}. The Type I Error rates are generated using the true source and sample mean as the test statistic $R_N^\text{obs}$ and the bootstrap distribution $\sqrt{m} \cdot d_{\mathcal{S}}([\bar q_{\boldsymbol{.},p}], [\bar q_{\boldsymbol{.},p}^*])$ as the null distribution across different percentiles $p\in\{0.65,0.75,0.85,0.95\}$, scale factors $m \in \{N^{0.6},N^{0.7},N^{0.8},N^{0.9},N\}$, and sample sizes $N \in \{ 50, 100, 500, 1000 \}$ with $\alpha=0.1$. The results here are not adjusted for multiple testing, and are instead based on separate, independent tests for each percentile to illustrate how Type I error varies across percentile level here.

As predicted from Figure~\ref{subfig:distances}, when the asymptotic scaling assuming a valid bootstrap procedure is used ($m=N$), the confidence intervals are conservative: for almost all percentile and sample size combinations, the Type I Error rate is 0 when it should be closer to $\alpha=0.1$. Interestingly, the only exception to this is the highest percentile $p=0.95$, and in fact using a lower asymptotic scale inflates the Type I error rate. This result is most likely due to the more smoothly varying gradient seen at higher percentiles of the sample mean images. The smoother the images are the more likely the function $c(f_p(\bar S))$ is to be differentiable.

At high sample sizes with non-smooth images we recommend using the $m$-out-of-$n$ bootstrap procedure. Based on these simulations for the ICF application, $m=N^{0.8}$ appears to work reasonably well as seen in Table~\ref{tab:type1error}. In the Supplementary Materials, we demonstrate how to select an $m$ by observing changes in the distribution of $\sqrt{m} \cdot d_{\mathcal{S}}([\bar q_{\boldsymbol{.},0.65}], [\bar q_{\boldsymbol{.},0.65}^*])$ for different values of $m$. When images are smooth or if there is any doubt about this procedure, setting $m=N$ is the safest option, as this results in conservative hypothesis tests. An example of smoother image data can be seen in the Supplementary Materials where the ESD of the true distribution $d_{\mathcal{S}}([q_p],[\bar q_p])$ and the bootstrap distribution $d_{\mathcal{S}}([\bar q_{\boldsymbol{.},p}],[\bar q^*_{\boldsymbol{.},p}])$ (Figure~\ref{subfig:distances_pc}) have less bias than the ones seen in Figure~\ref{subfig:distances}.

\subsection{Simulation: Power and Sensitivity Analysis}\label{subsec:sim2}

\begin{figure}
\centering
    \begin{subfigure}{0.29\linewidth}
    \centering
    \includegraphics[width=\linewidth]{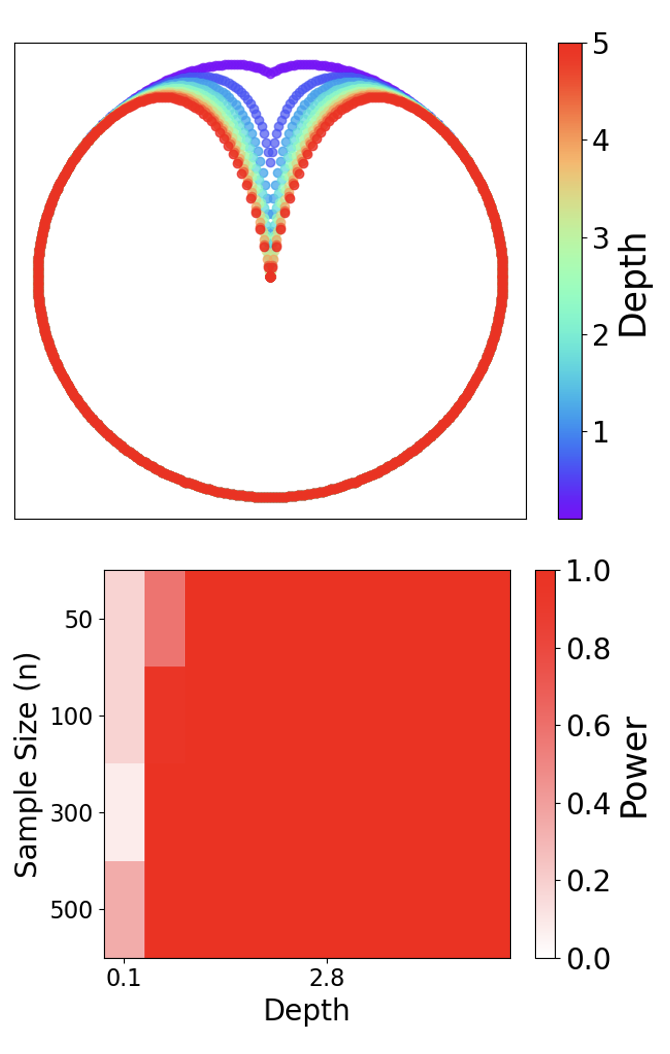}
    \caption{Power indented circles}\label{subfig:heartpower}
    \end{subfigure}
        \begin{subfigure}{0.3\linewidth}
    \centering
    \includegraphics[width=\linewidth]{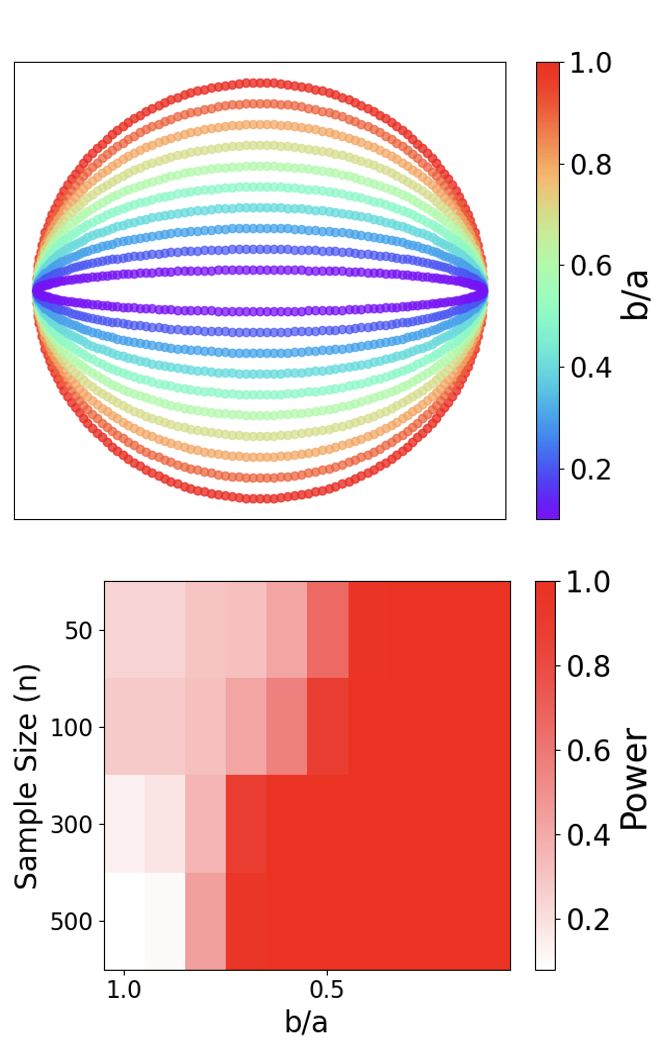}
    \caption{Power ellipses}\label{subfig:ellipsepower}
    \end{subfigure}
        \begin{subfigure}{0.32\linewidth}
    \centering
    \includegraphics[width=\linewidth]{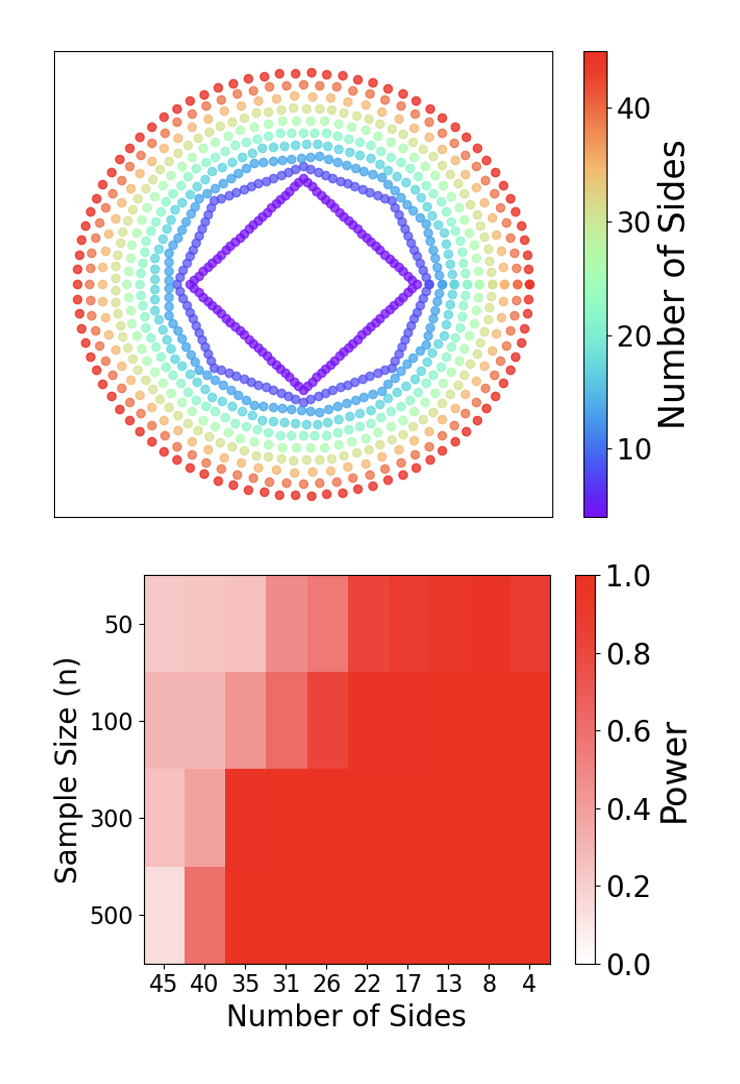}
    \caption{Power polygons}\label{subfig:polypower}
    \end{subfigure}
\caption{Power analysis of the hypothesis test when $c_p$ is a circle and $c^0$ is a (a) indented circle, (b) ellipse, or (c) polygon. The color in each plot indicates the parameter value that governs the shape such as (a) depth of indentation in the circle, (b) amount of difference in ellipse parameters $\frac{b}{a}$, or (c) number of sides of the polygon. The results of the power analysis are shown in the second row with a heat map.} \label{fig:simulationresults_sensitivity}
\end{figure}

This section examines the power of the hypothesis test, which measures the probability of correctly identifying substantive shape deviations away from a null shape. For this simulation the same data as described in Section~\ref{subsec:sim1} is used for the hypothesis test, but the test statistic $R_N^\text{obs}$ is different. We focus our attention on power analysis for shape deviations (from a circle) that are consistently observed in ICF images: indented circles (Figure~\ref{subfig:heartpower}), ellipses (Figure~\ref{subfig:ellipsepower}), or polygons (Figure~\ref{subfig:polypower}).

The primary question for this numerical study is whether the method can distinguish between different departures relative to a circle, and under what conditions such discrimination is possible. Starting with the shape of an indented circle as $c_p^0$ (e.g. this could be a fill-tube jet in ICF images), different depths $d\in\{0.1, 0.64, \ldots, 4.46, 5\}$ are tested to see how often the null hypothesis is rejected. For each depth and sample size $N\in\{50,100,300,500\}$, fifty hypothesis test are performed, and power is the percentage of times the test is rejected. The top image in Figure~\ref{subfig:heartpower} illustrates the different $c_p^0(d)$ where color corresponds to depth $d$. The bottom image in Figure~\ref{subfig:heartpower} shows the power across different sample sizes and depths. A power of one indicates that all fifty hypothesis tests rejected the null hypothesis that an indented circle of depth $d$ is the contour of the true source based on data generated from a true source contour of a circle. We use a scale of $m=N^{0.8}$ based on Section~\ref{subsec:sim1} since we are only testing data at the $65^\text{th}$ percentile. In general, only small depths with low sample sizes had a power less than one and it was still a very high power of $90\%$. 

For the ellipses in Figure~\ref{subfig:ellipsepower}, we systematically vary the ratio of its semi-major axis $a$ relative to semi-minor axis $b$ by shrinking $b/a$ from 1 (a circle) to 0.1 (a flat vertical ellipse). In ICF applications, this corresponds to examining the many implosion shots which are ellipsoidal (e.g., `pancake' or `hot dog' shape) in nature because  the fuel target is not compressed symmetrically. As seen in Figure~\ref{subfig:ellipsepower}, compared to the indented circles example, our test is less powerful in detecting changes in $c^0_{p}=c^0_p(b/a)$. For $b/a$ near one, a larger sample size is necessary for the test to have power close to one.

Finally, we look at polygons (Figure~\ref{subfig:polypower}) with varying numbers of sides $\{4,  8, \ldots, 40, 45 \}$ as the null shape $c^0_{p}=c^0_p(\text{sides})$. As expected, the hypothesis test for polygons with a low number of sides has a very high power even at low sample sizes. In general, the test has good power even with low sample sizes and the slight upwards bias in the approximate distribution seen in Section~\ref{subsec:sim1}.

\section{Real Data}\label{sec:data}

\begin{figure}
\centering
    \begin{subfigure}{0.23\linewidth}
    \centering
    \includegraphics[width=\linewidth]{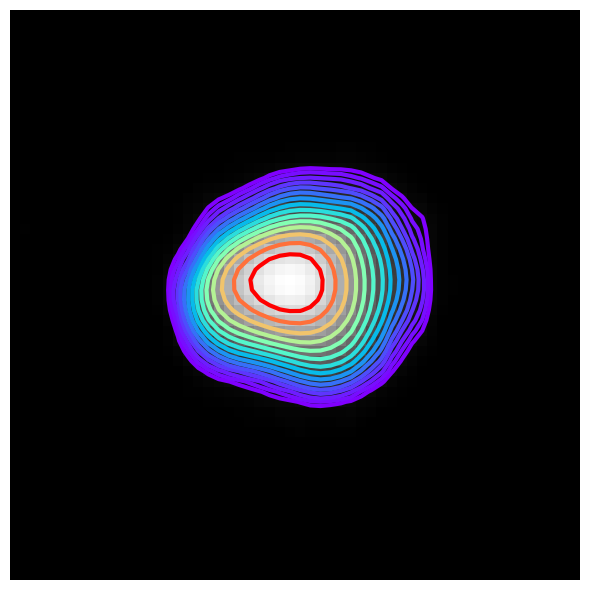}
    \caption{Sample mean $\bar S_1$}\label{subfig:data1}
    \end{subfigure}
    \begin{subfigure}{0.23\linewidth}
    \centering
    \includegraphics[width=\linewidth]{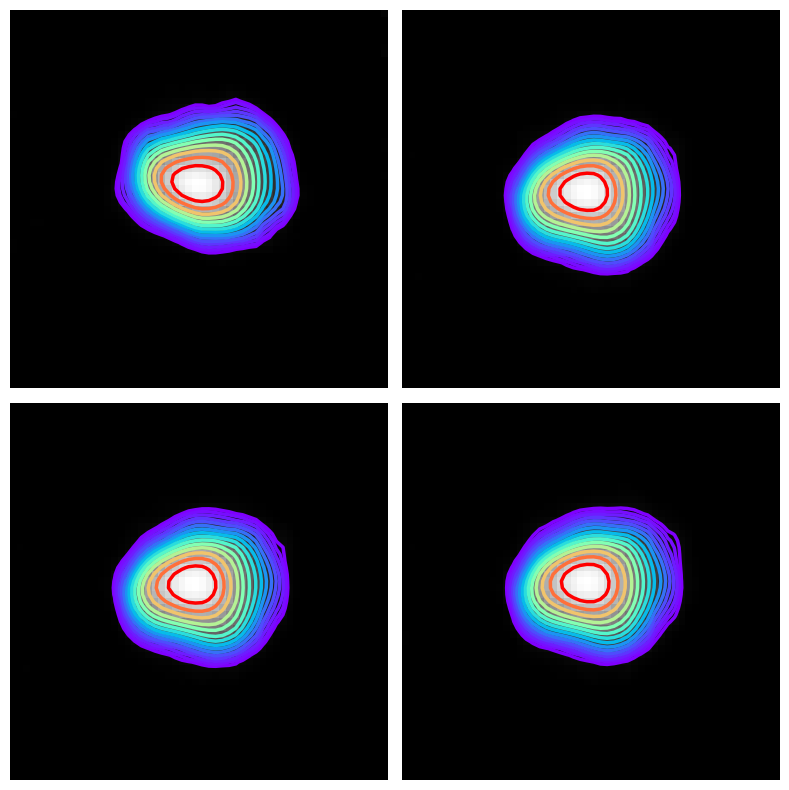}
    \caption{Bootstrapped $\bar S^*_1$}\label{subfig:bootsamp1}
    \end{subfigure}
    \begin{subfigure}{0.23\linewidth}
    \centering
    \includegraphics[width=\linewidth]{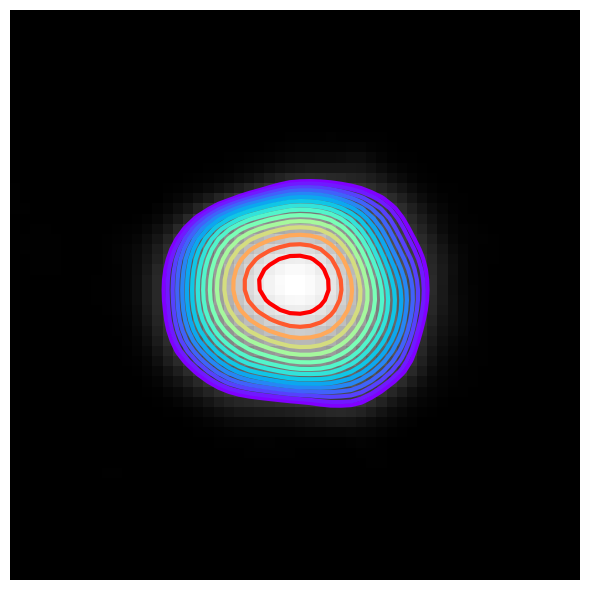}
    \caption{Sample mean $\bar S_2$}\label{subfig:data2}
    \end{subfigure}
    \begin{subfigure}{0.23\linewidth}
    \centering
    \includegraphics[width=\linewidth]{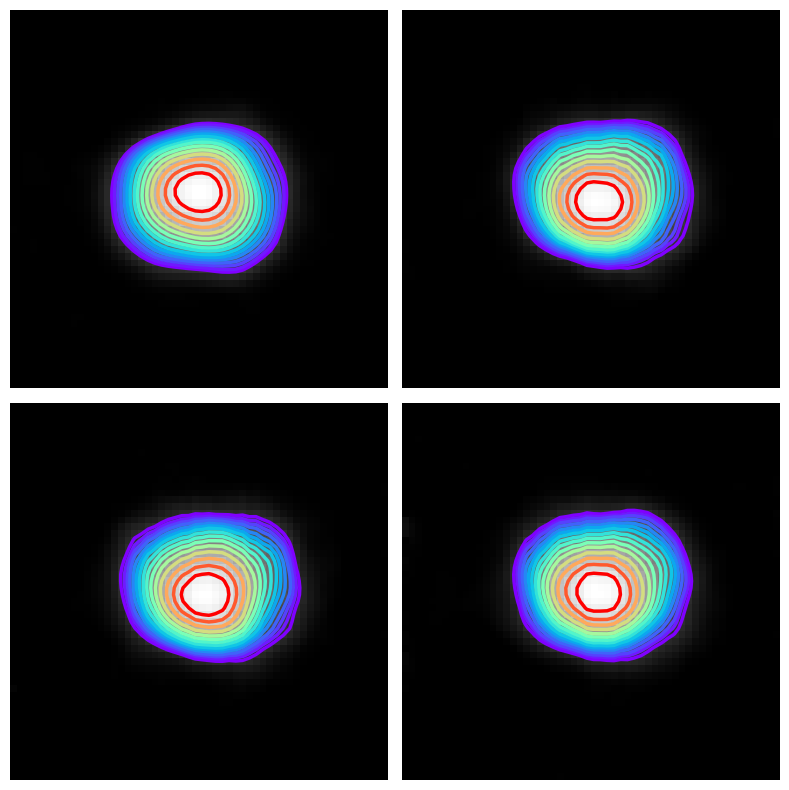}
    \caption{Bootstrapped $\bar S^*_2$}\label{subfig:bootsamp2}
    \end{subfigure}
\caption{Sample mean images (a) $\bar S_1$ and (c) $\bar S_2$ for two separate ICF shots. The contour color corresponds to percentile for $p=\{0.85,\ldots,1\}$ where purple is for the $85^\text{th}$ percentile and red is for the $100^\text{th}$ percentile. Four different bootstrap sample mean images are shown for (b) $\bar S_1^*$ and (d) $\bar S_2^*$ with corresponding percentile contours.} \label{fig:realdata}
\end{figure}

The {\bf bootESA} method is tested on two separate example ICF images as seen in Figure~\ref{subfig:data1} and Figure~\ref{subfig:data2} where the color of the contour corresponds to percentiles $p=\{0.85,0.86,\ldots,0.99,1\}$. The number of pinholes for both real-world examples is $N=55$: this includes pinhole sub-images with bad diagnostics for a reconstructed image (i.e., residuals not correct). In practice, only a subset of pinhole sub-images are used; however, we use all 55 pinhole reconstructed images in our analysis for demonstration purposes. Since these images are relatively smooth, even though the sample size is relatively small, we use $m=N$ as the asymptotic scaling to be conservative. Four bootstrapped sample means are shown for both images in Figure~\ref{subfig:bootsamp1} and Figure~\ref{subfig:bootsamp2} to illustrate variability in the estimates. The number of bootstrap sample mean images used to generate the empirical distribution (Section~\ref{subsub:step2}) for the hypothesis test is $B=1000$. 

The bootESA hypothesis testing procedure is performed on each percentile contour separately to assess whether each percentile's shape is a circle; we do this to illustrate how the test changes as the pre-selected percentile level changes. Since a single level is chosen a priori in practice, we do not consider adjusting for multiple testing across $p$ in this manuscript, leaving this for future work. The null hypothesis is that the true source image contour is a circle (though any null shape may be used), and we specify significance level $\alpha=0.05$.
\begin{equation}
\begin{split}
    H_0:[q^0] \in C_{\alpha,n} \text{ (i.e. }c_p \text{ is a circle)}  \\
    H_A:[q^0] \notin C_{\alpha,n} \text{ (i.e. }c_p \text{ is not a circle)}
\end{split}
\end{equation}

\begin{figure}
\centering
\begin{subfigure}{\linewidth}
    \centering
    \includegraphics[width=\linewidth]{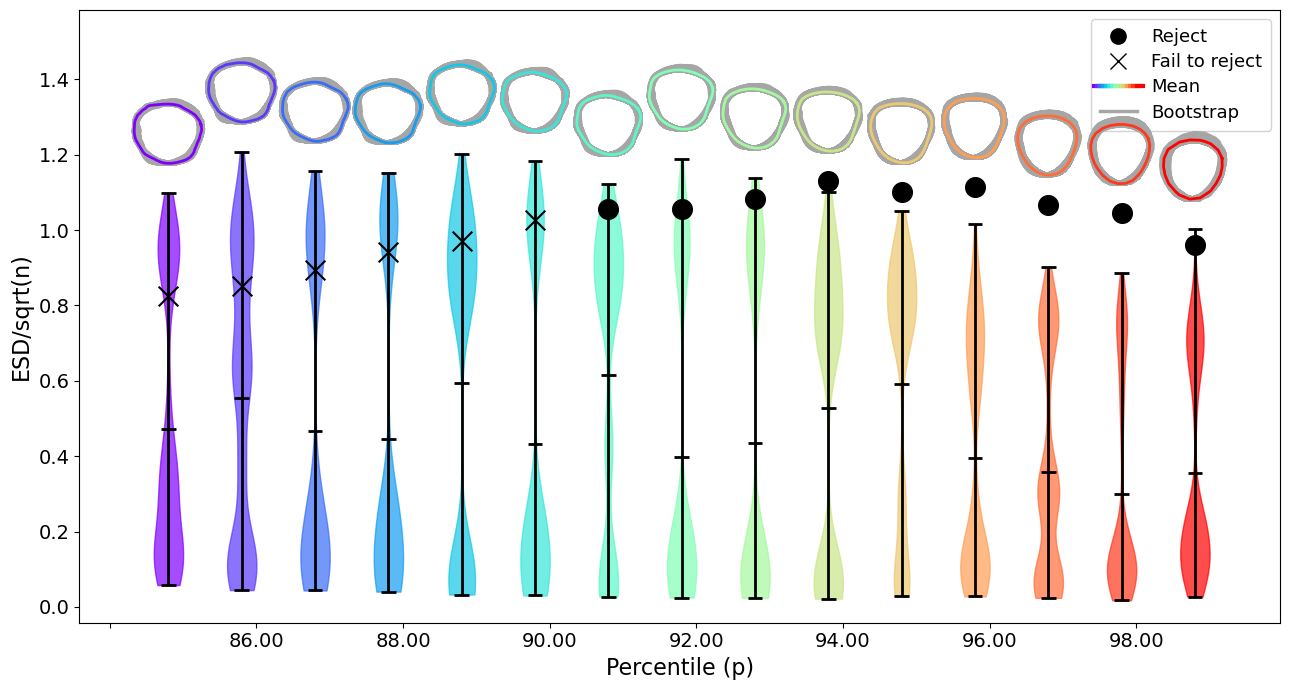}
    \caption{Hypothesis Test $\bar S_1$}\label{subfig:hyptest1}
    \end{subfigure}
    \begin{subfigure}{\linewidth}
    \centering
    \includegraphics[width=\linewidth]{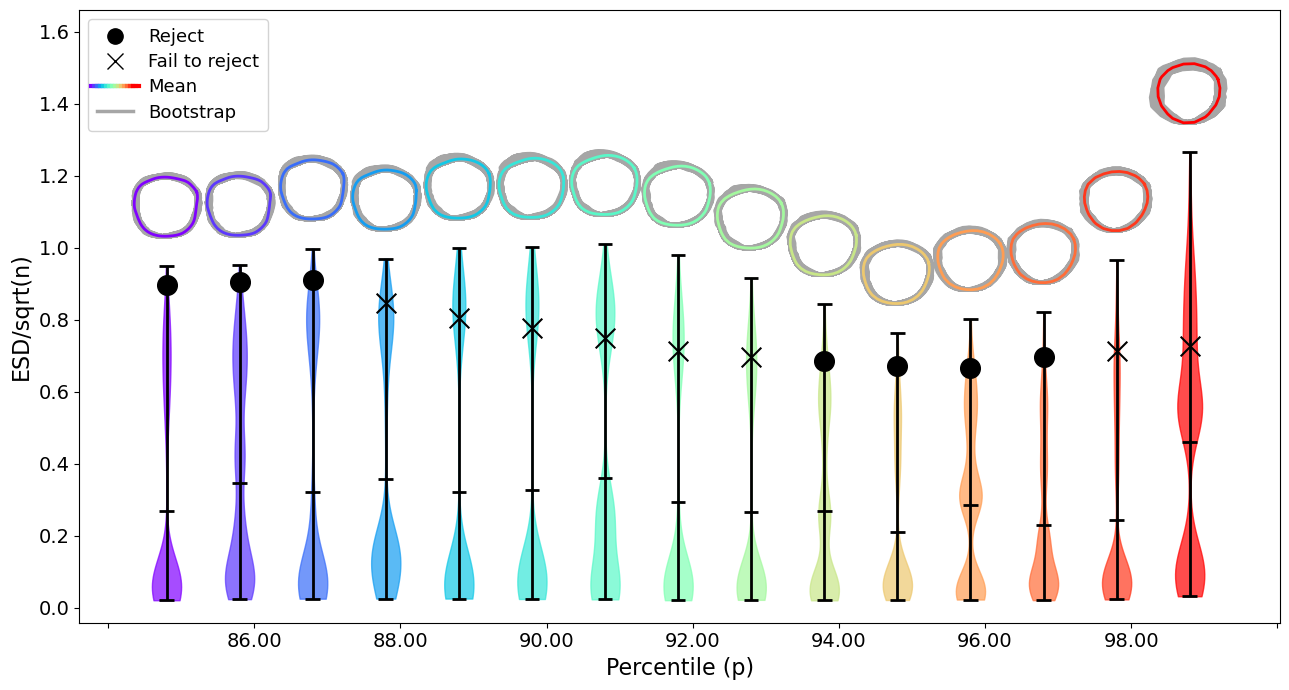}
    \caption{Hypothesis Test $\bar S_2$}\label{subfig:hyptest2}
    \end{subfigure}
\caption{Hypothesis test results for two different ICF implosions shots in (a) $\bar S_1$ and (b) $\bar S_2$. The contour color and $x-$axis values are the percentile level $p = \{ 0.85, 0.76, 0.77, \ldots, 0.98, 0.99 \}$. The test statistic is a $X$ for Fail to reject and a dot for Reject for each contour. Above the violin plots are $\bar c_p$ (color) and $\bar c^*_p$ (gray).} \label{fig:realdata_hypothesistest}
\end{figure}

The results of the hypothesis test by image and percentile are show in Figure~\ref{fig:realdata_hypothesistest}. The $x-$axis is the percentile level $p$ corresponding to the images in Figure~\ref{fig:realdata} where the contour colors are matched. The violin plots show the empirical distribution of distances for $d_{\mathcal{S}}([\bar q_1], [\bar q_1^*])$ and $d_{\mathcal{S}}([\bar q_2], [\bar q_2^*])$ for each percentile. The shape at the top of each violin plot is the contour of the sample mean image with the bootstrapped contours for that percentile underneath in gray. For each percentile contour the test statistic $R_N^{\text{obs}}$ is a point with either (i) a cross symbol if the hypothesis test result was Fail to reject or (ii) a dot if the hypothesis test result was Reject.

Both of the ICF shots have contours which are found to significantly differ from a circle and contours which are found to not be statistically significant (i.e. not significantly differ from a circle). The first shot (Figure~\ref{subfig:data1}) has more variation in lower percent contours ($p=\{0.85,\ldots,0.94\}$) and smaller effect sizes. Therefore only the first six percentiles have contours that are close enough to a circle with large enough variability that the result is Fail to Reject the null. The second shot (Figure~\ref{subfig:data2}) has the most variability for the higher percentiles and low effect sizes for $p=\{0.9,\ldots,1\}$. The middle five percentile contours and the highest two percent contours are found to not be statistically different from a circle. 

In general, the results of the hypothesis test are consistent with intuition. Though the hypothesis test is only created to test one image at a time, a potential interpretation could be that the second shot is closer to a circle. This is because the second shot has more hypothesis test results of Fail to reject, the effect sizes are generally smaller and there is less variability (except for the highest two percentiles).

\section{Conclusion and Discussion}

In this work, we present a new statistical inference approach to analyzing the shapes of objects in images using ESA. This approach advances the ESA literature by creating a new approximate hypothesis testing framework for one-sample tests of shape. Following ideas in other areas of shape analysis such as topological data analysis and Procrustes-based methods, we extend confidence intervals generated from a bootstrap empirical distribution to ESA analysis of contours in an image. We demonstrate the methods' performance (via Type I error rates) along with the power and sensitivity of the hypothesis test through several numerical studies. Then, we demonstrated the proposed procedure on real-world data examples in ICF experiments to try and understand the shape of the source.  This approach extends well beyond ICF images, allowing for the estimating of the underlying structure of shapes in a data space as long as there is a valid way to bootstrap the space.

While this method is very powerful and useful in its own right, ICF applications in particular are interested in the symmetry of the source. One area of extension would be to develop a procedure which measures degree of symmetry or location of asymmetry. Our method also assumes that the topology of each contours we are comparing is the same. Another interesting extension to the method would be trying to understand how to extend the test to accommodate contours with different topologies. Furthermore, our method treats each contour independently across percentiles; in reality, there is a natural hierarchical structure to contours across percentiles, that we could exploit to test all contours simultaneously (avoiding need to pre-select a percentile for analysis) or identifying the contours that best describe the shape of the underlying structure. Based on the Type I error analysis, the contours do not necessarily behave the same across percentiles, with some levels having increased variability. Lastly, an interesting extension would be to treating the image as a surface and use analogous methods to compare three-dimensional shapes.


\section{Data Availability Statement}\label{data-availability-statement}

The simulated data and code has been made available at the following URL: \url{https://github.com/ssglenn-LANL/bootESA}. 

The data that support the findings of this study are available from the corresponding author, CD, upon reasonable request.

\bigskip
\begin{center}
{\large\bf SUPPLEMENTAL MATERIALS}
\end{center}

\begin{description}

\item[Other Data Generating Example:]\label{sup:KDE}
As stated in Section~\ref{sec:background}, while we present {\bf bootESA} in terms of ICF image data, this method generalizes to any problem with a space of interest $S$, some learned mean of that space $\bar S$, and bootstrapped resamples of the learned mean $\bar S^*$. One such example is a typical dataset used in Topological Data Analysis (TDA). Let the data $\bf{Y}$ be a random sample of points on a manifold $\mathcal{M}$ from some kernel density $S=\rho$ where $Y^1, \ldots, Y^N \overset{iid}{\sim} \rho$. The estimate of $\rho$ is a kernel density estimate (KDE) of the points $\bar S = \hat \rho$. 

In other words, we want to estimate the shape of the manifold the data was generated by estimating the shape of $\rho$ with $\hat \rho$. The KDE is a consistent estimate of the true kernel density conditional on the correct bandwidth parameter. As the points are assumed to be sampled $iid$ from $\rho$ we can bootstrap $\bf{Y}$ to generate $\bf{Y}^*$ and $\bar S^*=\hat \rho^*$. 

Following the {\bf bootESA} method in Section~\ref{subsec:bootESAmethod}, a contour is used to represent the shape of the manifold (i.e. the density points are drawn from) denoted by $\bar c_p$ and bootstrapped densities $\bar c^*_p$. An example of this procedure can be seen in Figure~\ref{fig:simulation_pointcloud} where $\rho$ is in Figure~\ref{subfig:rho}, examples of $\hat \rho$ are in Figure~\ref{subfig:hatrho}, and examples of $\hat \rho^*$ are in  Figure~\ref{subfig:hatrhostar} with corresponding contours $c_{0.6}, \bar c_{0.6}, \bar c_{0.6}^*$ in red.

\begin{figure} 
\centering
    \begin{subfigure}{0.3\linewidth}
    \centering
    \includegraphics[width=\linewidth]{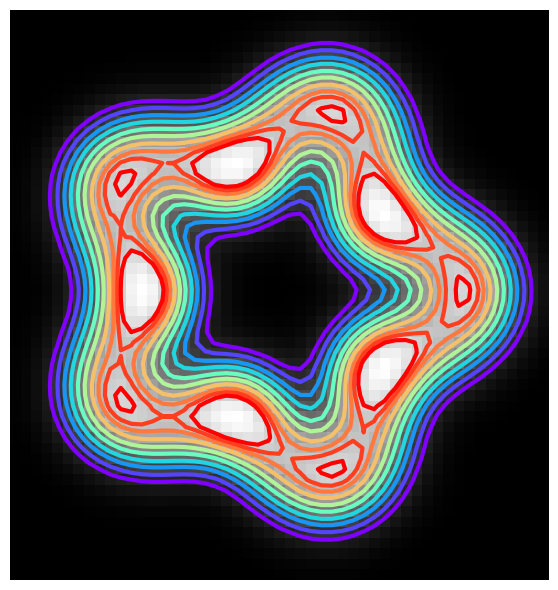}
    \caption{Density $\rho$}\label{subfig:rho}
    \end{subfigure}
    \begin{subfigure}{0.3\linewidth}
    \centering
    \includegraphics[width=\linewidth]{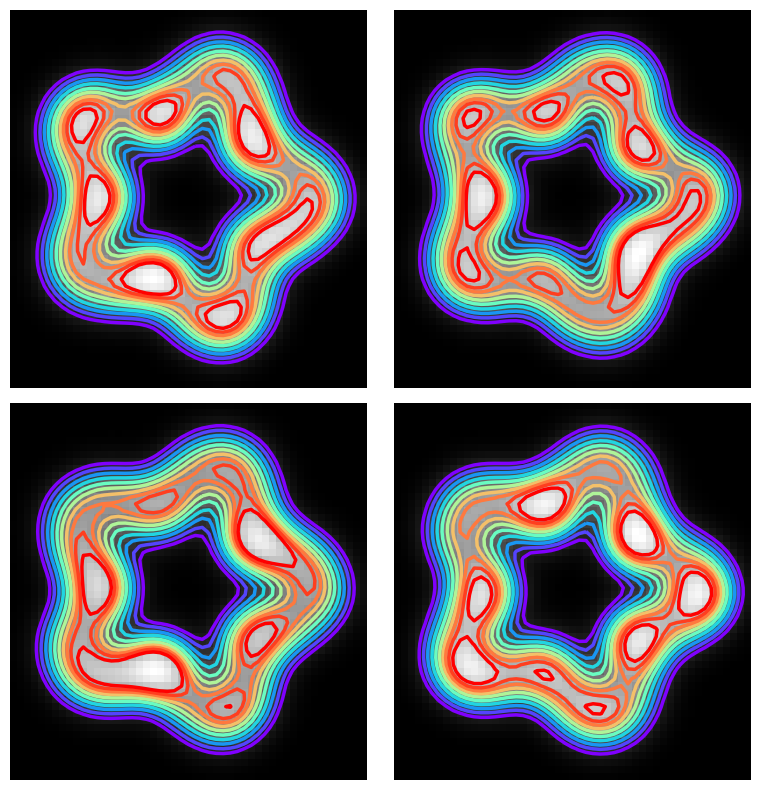}
    \caption{KDE $\hat \rho$}\label{subfig:hatrho}
    \end{subfigure}
    \begin{subfigure}{0.3\linewidth}
    \centering
    \includegraphics[width=\linewidth]{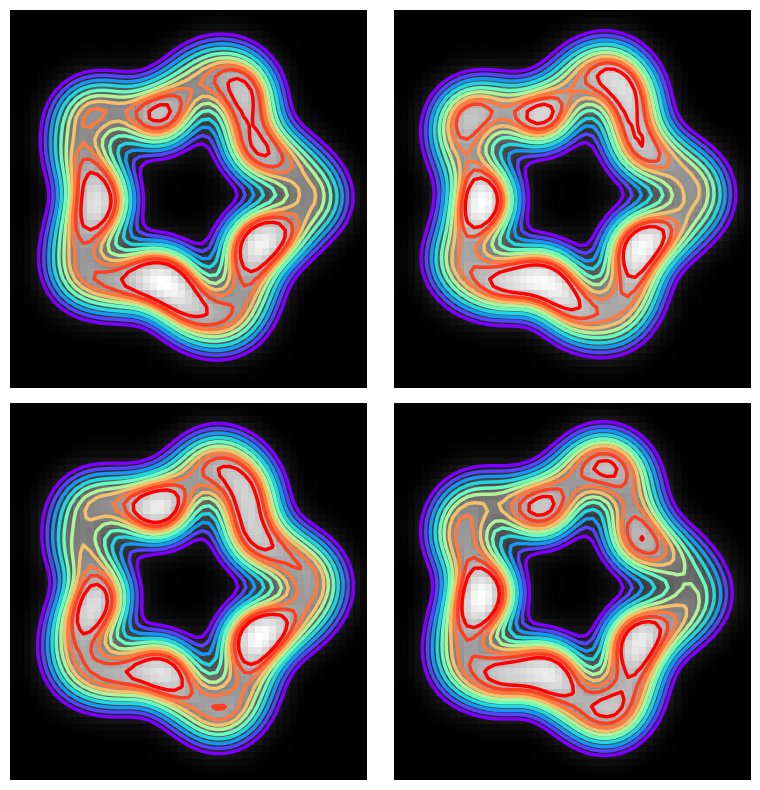}
    \caption{Bootstrap KDE $\hat \rho^*$}\label{subfig:hatrhostar}
    \end{subfigure}
\caption{Example of the data setup with (a) the true density, (b) data estimates using the KDE, and (c) bootstrapped KDEs with $60^\text{th}$ percentile contours (red).} \label{fig:simulation_pointcloud}
\end{figure}

A similar type of numerical study as in Section~\ref{subsec:sim1} is carried out but sample size $N$ is points in $\mathbb{R}^2$ instead of pinhole images, and the data estimate is a KDE not a sample mean image. The number of points sampled is $N =\{300, 500, 1000, 6000, 10000\}$ with 100 observations per sample size group for the distribution of ESD. The red is the approximate distribution $d_{\mathcal{S}}([\bar q_{0.6}], [\bar q_{0.6}^*])$ and the blue is the true distribution $d_{\mathcal{S}}([q_{0.6}], [\bar q_{0.6}])$ in Figure~\ref{fig:simulationresults_pc}. 

We used a percentile contour of $p=0.6$ which was the largest $p-$level where the topology of the contour did not change for small perturbations in the data, a fixed, closed curve even for small $N$. For example at low sample sizes a $0.6$ percentile contours sometimes lead to a C shape contour as opposed to a fully connected flower shaped contour. While this was seen in both the bootstrap and true ESD distributions, this fact would violate Assumption~\ref{as:topology} and for visualization purposes we wanted to keep the contours at a fixed topology. As seen in this example, the approximate distribution and the true distribution of ESD behave similarly as in Section~\ref{subsec:sim1}, though the data is completely different. In this example, there appears to be less noise which in part is due to the ability to smooth the data using the bandwidth parameter. In general, we show the robustness and generalizability of the {\bf bootESA} method with this simulation.
\begin{figure}
\centering
    \begin{subfigure}{0.3\linewidth}
    \centering
    \includegraphics[width=\linewidth]{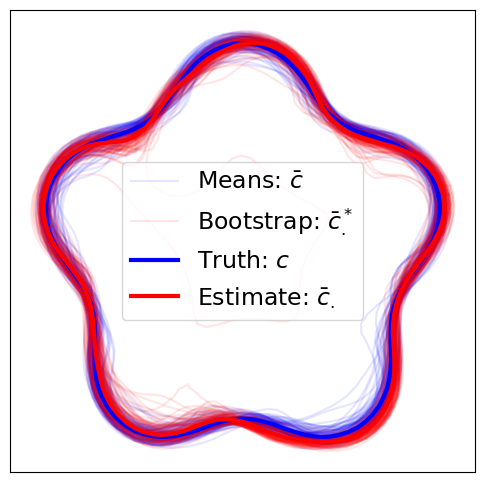}
    \caption{$\bar c_{0.6}, \bar c_{0.6}^*$ for $N=200$}\label{subfig:meancountours1_pc}
    \end{subfigure}
    \begin{subfigure}{0.3\linewidth}
    \centering
    \includegraphics[width=\linewidth]{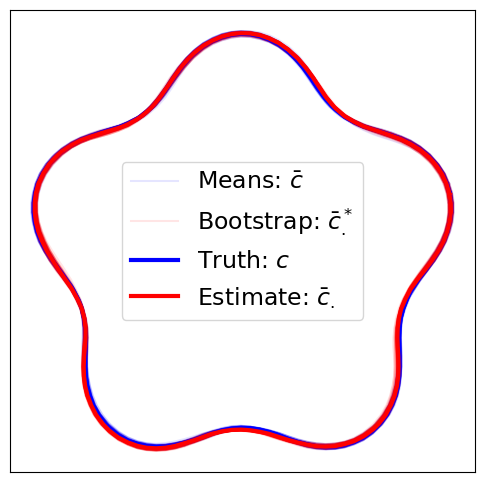}
    \caption{$\bar c_{0.6}, \bar c_{0.6}^*$ for $N=5000$}\label{subfig:meancountours2_pc}
    \end{subfigure}
    \begin{subfigure}{0.28\linewidth}
    \centering
    \includegraphics[width=\linewidth]{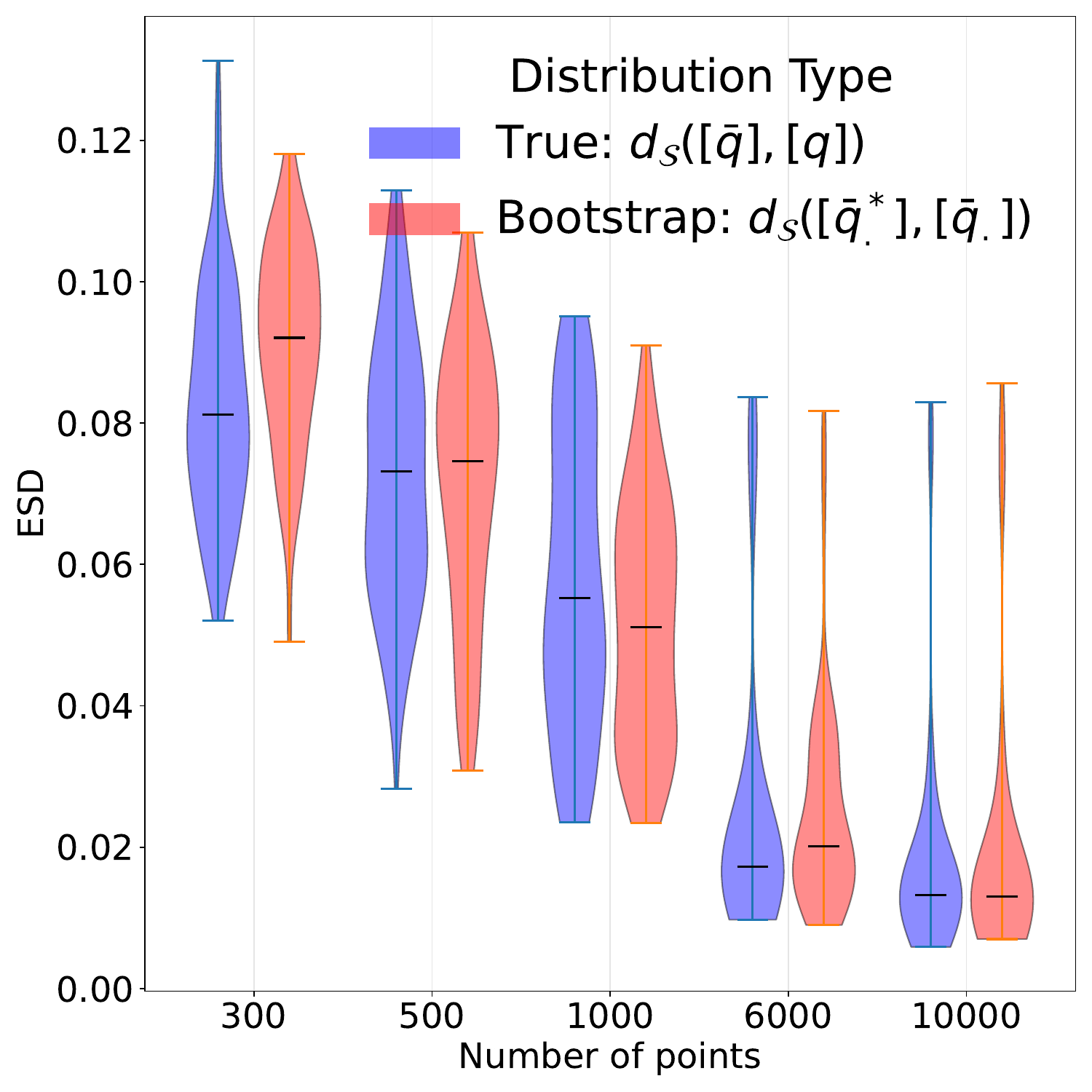}
    \caption{Violin plots ESD}\label{subfig:distances_pc}
    \end{subfigure}
\caption{Simulation results for sample mean contours $\bar c_{0.6}$ and the source contour $c_{0.6}$ are in blue; the contour of the mean $\bar c_{0.6}$ and bootstrap contours $\bar c_{0.6}^*$ are in red with sample sizes of (a) $200$ points and (b) $5000$ points. (c) The distribution of raw ESDs for the true empirical distribution (blue) and the approximate empirical distribution (red) for sample size (x-axis).}\label{fig:simulationresults_pc}
\end{figure}

\item[Simulations for $m$ out of $n$ bootstrap:] \label{sup:moutofn}

\begin{figure}
\centering
\begin{subfigure}{\linewidth}
    \centering
    \includegraphics[width=0.7\linewidth]{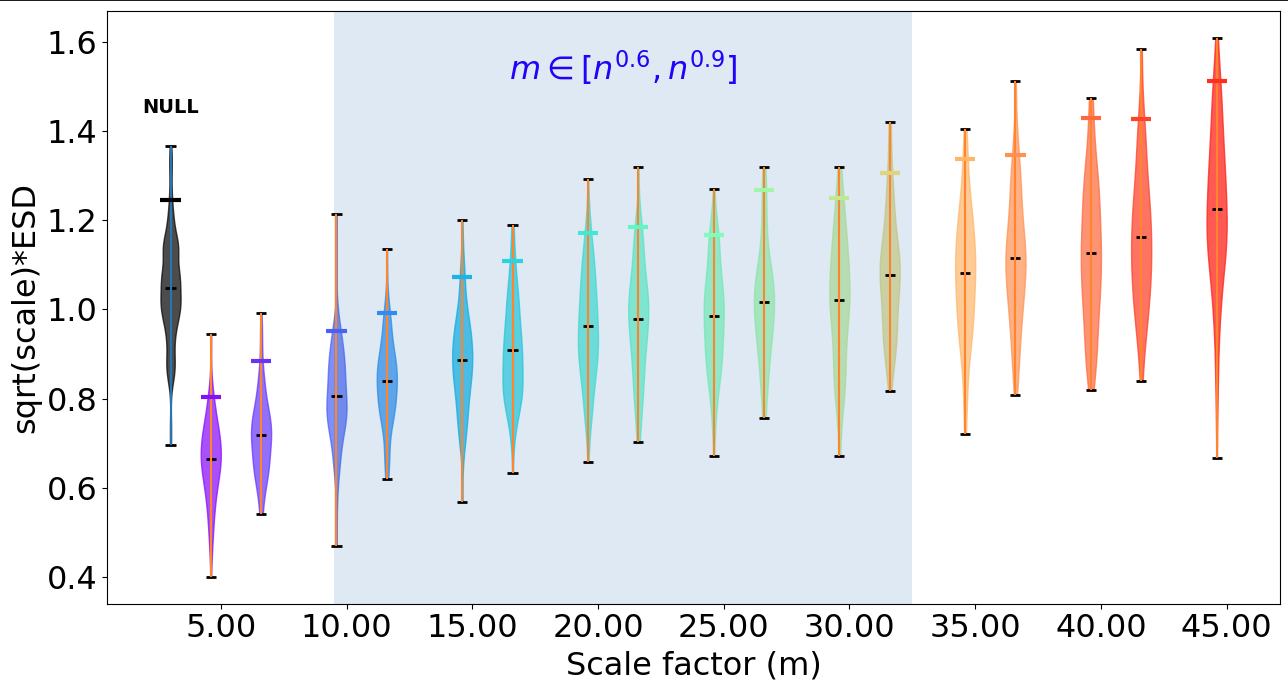}
    \caption{$N=50$}\label{subfig:n}
    \end{subfigure}
    \begin{subfigure}{\linewidth}
    \centering
    \includegraphics[width=0.7\linewidth]{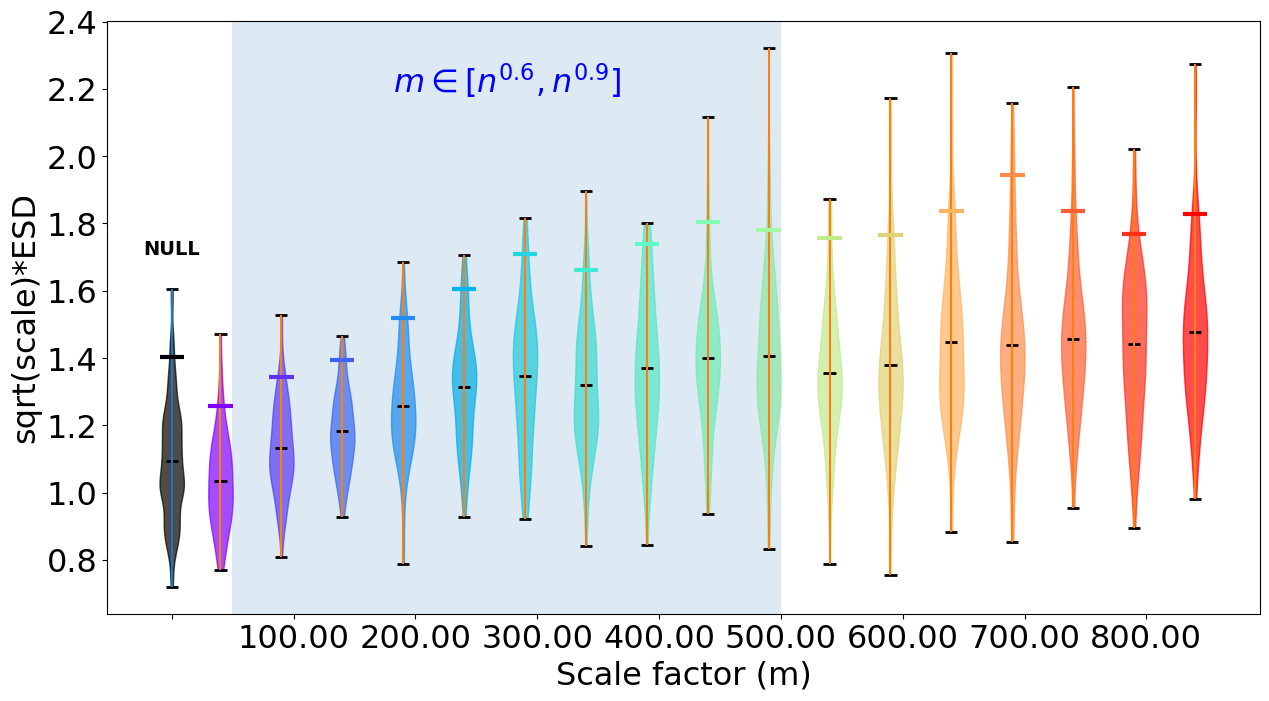}
    \caption{$N=1000$}\label{subfig:n}
    \end{subfigure}
\caption{Simulations for finding stability for $m$ in bootstrap with the target distribution (black) and $m \in \{N^\epsilon\}$ on the $x-$axis for $\epsilon \in [0.5,1]$.} \label{fig:moutofn}
\end{figure}

In this section we show the practice of considering a range of $m$ values, increasing $m$ until a chosen distance between successive distributions becomes small and remains stable. These simulations are done on the same data as shown in Section~\ref{subsec:sim1} to get the range of $\epsilon$ values in Table~\ref{tab:type1error}. We again focus on $p=0.65$ for simplicity and look at how the distribution $d_{\mathcal{S}}([\bar q_{\boldsymbol{.}0.65}], [\bar q^*_{\boldsymbol{.}0.65}])$ converges to the true empirical distribution $d_{\mathcal{S}}([q_{0.65}], [\bar q_{0.65}])$ with $B=1000$ bootstrap samples for sample sizes $N=50 $ and $N=1000$. A typical range for $\epsilon \in [0.6,0.9]$ with $0.9$ being conservative and behaving well for small sample sizes and smoother images and $0.6$ working well when the bootstrap procedure fails (i.e. high sample sizes and less smooth images). This range is highlighted in blue in Figure~\ref{fig:moutofn} and the true empirical distribution in black. The colors of each violin plot is based on sample size with red being the highest $m$ and purple being the smallest $m$.

\end{description}

{}



\end{document}